\documentclass{aa}
\usepackage{graphicx}
\usepackage{txfonts}

\usepackage[english]{babel}

\usepackage{natbib}
\bibpunct{(}{)}{;}{a}{}{,}

\renewcommand{\imath}{\mathrm{i}}

\begin{document} 

  \title{Constraining regular and turbulent magnetic field strengths in M51 via Faraday depolarization}
  \titlerunning{Constraining magnetic field strengths in M51}
   \author{C. Shneider \inst{1} \and M. Haverkorn \inst{$2,1$} \and A. Fletcher \inst{3} \and A. Shukurov \inst{3} }

    \institute{Leiden Observatory, Leiden University,P.O. Box 9513, 2300 RA Leiden, The Netherlands \\ \email{shneider@strw.leidenuniv.nl} \and  Department of Astrophysics/IMAPP, Radboud University Nijmegen, P.O. Box 9010, 6500 GL Nijmegen, The Netherlands \and School of Mathematics and Statistics, Newcastle University, Newcastle upon Tyne NE1 7RU, U.K.}

   \date{Accepted 18 June 2014} 

  \abstract{We employ an analytical model that incorporates both wavelength-dependent and wavelength-independent depolarization to describe radio polarimetric observations of polarization at $\lambda \lambda \lambda \, 3.5, 6.2, 20.5$ cm in M51 (NGC 5194). The aim is to constrain both the regular and turbulent magnetic field strengths in the disk and halo, modeled as a two- or three-layer magneto-ionic medium, via differential Faraday rotation and internal Faraday dispersion, along with wavelength-independent depolarization arising from turbulent magnetic fields. A reduced chi-squared analysis is used for the statistical comparison of predicted to observed polarization maps to determine the best-fit magnetic field configuration at each of four radial rings spanning $2.4 - 7.2$ kpc in $1.2$ kpc increments. We find that a two-layer modeling approach provides a better fit to the observations than a three-layer model, where the near and far sides of the halo are taken to be identical, although the resulting best-fit magnetic field strengths are comparable. This implies that all of the signal from the far halo is depolarized at these wavelengths. We find a total magnetic field in the disk of approximately $18~\mu$G and a total magnetic field strength in the halo of $\sim 4-6~\mu$G. 
Both turbulent and regular magnetic field strengths in the disk exceed those in the halo by a factor of a few. About half of the turbulent magnetic field in the disk is anisotropic, but in the halo all turbulence is only isotropic.}


   \maketitle

\section{Introduction} 

   Magnetic fields are important drivers of dynamical processes in the interstellar medium (ISM) of galaxies on both large and small scales. 
   They regulate the density and distribution of cosmic rays in the ISM \citep{beck04} and couple with both charged and, through ion-neutral collisions, neutral particles in essentially all interstellar regions except for the densest parts of molecular clouds \citep{ferriere}. 
   Moreover, their energy densities are comparable to the thermal and turbulent gas energy densities on large scales, as indicated for the spiral galaxies NGC 6946 and M33 and for the Milky Way \citep{beck07, taba08, marijke_heiles}, thereby affecting star formation and the flow of gas in spiral arms and around bars \citep[and refs. therein]{beck09_spacesci, beck07}.
   In the case of the Galaxy, magnetic fields contribute to the hydrostatic balance and stability of the ISM on large scales, while they affect the turbulent motions of supernova remnants and superbubbles on small scales \citep[and refs. therein]{ferriere}.
   Knowledge of the strength and structure of magnetic fields is therefore paramount to understanding ISM physics in galaxies. 

   Multiwavelength radio-polarimetric observations of diffuse synchrotron emission in conjunction with numerical modeling is a way of probing magnetic field interactions with cosmic rays and the diffuse ISM in galaxies. 
   Of particular interest are the total magnetic field and its regular and turbulent components, as well as their respective contributions to both wavelength-dependent and wavelength-independent depolarization in the thin and thick gaseous disk (hereafter the disk and halo). 

   Physically, regular magnetic fields are produced by dynamo action, by anisotropic random fields from compression and shearing gas flows, and by isotropic random fields by supernovae and other sources of turbulent gas flows. 
   In the presence of magnetic fields, cosmic ray electrons emit linearly polarized synchrotron radiation. 
Polarization is attributable only to the ordered magnetic fields, while unpolarized synchrotron radiation stems from disordered magnetic fields. 
   The degree of polarization $p$, defined as the ratio of polarized synchrotron to total synchrotron intensity, thus characterizes the magnetic field content and may be used as an effective modeling constraint.
   
 Except for edge-on galaxies, where the disk and halo are spatially distinct in projection to the observer, disentangling contributions to depolarization from the disk and halo is challenging. In this paper, we apply the theoretical framework developed in \citet{shneiderI} to numerically simulate the combined action of depolarization mechanisms in two or three consecutive layers describing a galaxy's disk and halo to constrain the regular and turbulent disk and halo magnetic field strengths in a face-on galaxy.
   
 In particular, M51 (NGC 5194) is ideally suited to studying such interactions for several reasons: (i) small angle of inclination ($l = -20^{\circ}$) permits the assumption of a multilayer decomposition into disk and halo components along the line of sight, (ii) high galactic latitude (b = $+68.6^{\circ}$) facilitates polarized signal extraction from the total synchrotron intensity since the contribution from the Galactic foreground is negligible at those latitudes \citep{berk}, and (iii) proximity of 7.6 Mpc allows for a high spatial resolution study.
   Besides a regular, large-scale magnetic field component and an isotropic random, small-scale field, the presence of an anisotropic random field component is expected since there is no large-scale pattern in Faraday rotation accompanying M51's magnetic spiral pattern observed in radio polarization \citep{fletcher11}. 
   Additionally, M51's galaxy type (Sc), linear dimension, and ISM environment are comparable with that of the Milky Way \citep{mao}, (see also \citet{pavel} for near infrared (NIR) polarimetry), possibly allowing for the nature of the global magnetic field properties of our own Galaxy to be further elucidated.

\section{Observational data} \label{obs_data}

We use the \citet{fletcher11} $\lambda \lambda \lambda \, 3.5, 6.2, 20.5$ cm continuum polarized and total synchrotron intensity observations of M51, taken with the VLA and Effelsberg and smoothed with a $15''$ beam resolution, to construct degree of polarization $p$ maps. The $p$ maps are partitioned into four radial rings from $2.4 - 7.2$ kpc in $1.2$ kpc increments with every ring further subdivided into $18$ azimuthal sectors, each with an opening angle of $20^{\circ}$, following \citet{fletcher11}. 
We will call these rings 1 through 4 from the innermost to the outermost ring.
This results in a total of $72$ bins.  
In the outermost ring, two of the bins are excluded as the number of data points within them is too small (less than five). 
For each of the remaining bins, histograms are produced to check that the individual distributions are more or less Rician and the mean of $p$ is computed with the standard deviation of $p$ taken as the error. 
Thermal emission subtraction was done using a constant thermal emission fraction across the Galaxy \citep{fletcher11}. In this method, thermal emission may have possibly been underestimated in the spiral arms in the \citet{fletcher11} total synchrotron intensity maps, the values of $p$ may, consequently, be overestimated in the bins that contain the spiral arms.

\section{Model} \label{model}

\subsection{Regular field} \label{regmagfld}

 Following \citet{fletcher11}, we use a two dimensional regular magnetic field $\sum_m \boldsymbol{B}_m(r) \, \cos \left(m \, \phi - \beta_m\right)$ for both the disk and halo with integer mode number $m$ and azimuthal angle in the galaxy plane $\phi$, measured counterclockwise from the northern end of the major axis along M51's rotation. 
A superposition of axisymmetric modes ($m= 0,2$) describes the disk magnetic field while mainly a bisymmetric mode ($m = 1$) describes the halo magnetic field. 
These modes yield the individual amplitudes $B_m$, pitch angles\footnote{The pitch angle of the total horizontal magnetic field is given by $\arctan \left( B_r / B_{\phi} \right)$ per mode $m$. Hence, $\sin\left(p_m\right)$ and $\cos\left(p_m\right)$ correspond to the $B_r$ and $B_\phi$ components of $\boldsymbol{B}$, respectively.} $p_m$ and $\beta_m$ angles\footnote{The $\beta$ angle is the azimuth at which the corresponding $m \neq 0$ mode is a maximum.}. 

The regular disk and halo magnetic fields in cylindrical polar coordinates are 
 \begin{align} \label{fletcherregflds}
 B_r & = B_0 \sin(p_0) + B_2 \sin(p_2)\cos(2 \phi - \beta_2), \nonumber \\
 B_{\phi} & =  B_0 \cos(p_0) + B_2 \cos(p_2)\cos(2 \phi - \beta_2), \nonumber \\
 B_z & = 0, \nonumber \\
 B_{\text{h}r} & = B_{\text{h}0}\sin(p_{\text{h}0}) + B_{\text{h}1}\sin(p_{\text{h}1})\cos(\phi - \beta_{\text{h}1}), \nonumber \\
 B_{\text{h}\phi} & = B_{\text{h}0}\cos(p_{\text{h}0}) + B_{\text{h}1}\cos(p_{\text{h}1})\cos(\phi - \beta_{\text{h}1}), \nonumber \\
 B_{\text{h}z} & = 0,
 \end{align}
where $h$ denotes the component of the halo field. 
Please consult Table~\ref{fletcherparam} for the associated magnetic field parameters in Eq.~\eqref{fletcherregflds} and see Fig.14 of \citet{fletcher11} for an illustration of their best-fit disk and halo modes.  
An anomalous halo pitch angle of $-90^{\circ}$ for the outermost ring was deemed unphysical and probably arose owing to the low polarization degrees in this ring. 
Therefore, we ignore this value and instead use $-50^{\circ}$, the pitch angle in the adjacent ring. 

Our model inputs only the regular magnetic field \emph{directions}, described by the respective modes for the disk and halo in Eq.~\eqref{fletcherregflds}, along with the relative strengths of these modes, given by $B_2 / B_0$ and $B_{h1}/B_{h0}$ in Table~\ref{fletcherparam}, while the regular disk and halo magnetic field \emph{strengths} are allowed to vary. 

The components of the regular magnetic field are projected onto the sky-plane \citep{berk} as 
 \begin{align*} 
  \overline B_x & = B_r \cos(\phi) \, - \, B_{\phi} \sin(\phi),				\nonumber \\
  \overline B_y & = \left[ B_r \sin(\phi) \, + \, B_{\phi} \cos(\phi) \right] \cos(l) + B_z \sin(l), \nonumber \\
  \overline B_{\parallel} &=  - \left[ B_r \sin(\phi) \, + \, B_{\phi} \cos(\phi) \right] \sin(l) + B_z \cos(l), \nonumber \\	  
 \end{align*}
where $l$ is the inclination angle and $\parallel$ denotes a component of the field parallel to the line of sight.

\begin{table}
\caption{Fitted Model Parameters adopted from \citet[Table A1]{fletcher11}. 
Ratios of mode strengths are reported as this allows for the magnetic field strengths to be left as a variable parameter in our model.}
\label{fletcherparam}
\begin{center}
\begin{tabular}{ l  c  c  c  c }
\hline\hline
\rule{0pt}{3ex} 
& Ring 1 & Ring 2 & Ring 3 & Ring 4 \\ \hline
\rule{0pt}{2.5ex}
$r$ [kpc] & $[2.4,3.6]$ & $[3.6,4.8]$ & $[4.8,6.0]$ & $[6.0,7.2]$ \\ 
\rule{0pt}{2.5ex}
$B_{2}/B_{0}$ & ${-33}/{-46}$ & ${-25}/{-57}$ & ${-40}/{-76}$ & ${-44}/{-76}$	\\ 
\rule{0pt}{2.5ex}
$\textit{p}_{0} [{}^{\circ}]$ & $-20$ & $-24$ & $-22$ & $-18$ \\ 
\rule{0pt}{2.5ex}
$\textit{p}_{2} [{}^{\circ}]$ & $-12$ & $16$ & $8$ & $3$ \\ 
\rule{0pt}{2.5ex}
$\beta_{2} [{}^{\circ}]$ & $-8$ & $-6$ & $-14$ & $-25$ \\ 
\rule{0pt}{2.5ex}
$B_{h1}/B_{h0}$ & $76/23$ & $...$ & $...$ & $...$ \\ 
\rule{0pt}{2.5ex}
$\textit{p}_{h0} [{}^{\circ}]$ & $-43$ & $...$ & $...$ & $...$ \\
\rule{0pt}{2.5ex}
$\textit{p}_{h1} [{}^{\circ}]$ & $-45$ & $-49$ & $-50$ & $-50\tablefootmark{a}$ \\ 
\rule{0pt}{2.5ex}
$\beta_{h1} [{}^{\circ}]$ & $44$ & $30$ & $-3$ & $-16$ \\ 
\hline
\end{tabular} 
\tablefoot{
The index $h$ refers to the halo magnetic field.
Dots mean that the corresponding parameter was insignificant in the \citet{fletcher11} fits and is thus not an input in our model. \\
\tablefoottext{a}{changed from original value of $-90^{\circ}$ to be in closer agreement with the halo pitch angle value reported for inner three rings.} 
}
\end{center}
\end{table}

\subsection{Turbulent field} \label{turbumagfld}

We explicitly introduce three-dimensional turbulent magnetic fields with both isotropic and anisotropic components.
The random magnetic fields are expressed as the standard deviations of the total magnetic field and are given by 
 \begin{align} 
 \sigma^2_x &= \sigma^2_r \left[ \cos^2(\phi) + \alpha \sin^2(\phi) \right], \nonumber \\ 
 \sigma^2_y &= \sigma^2_r \left\{ \left[ \sin^2(\phi) + \alpha \cos^2(\phi) \right] \cos^2(l) + \sin^2(l) \right\}, \nonumber \\ 
 \sigma^2_{\parallel} &= \sigma^2_r \left\{ \left[ \sin^2(\phi) + \alpha \cos^2(\phi) \right] \sin^2(l) + \cos^2(l) \right\}. \label{sigma_eq}
 \end{align}
Anisotropy is assumed to exclusively arise from compression along spiral arms and by shear from differential rotation and is assumed to have the form $\sigma^2_{\phi} = \alpha \, \sigma^2_r$ with $\alpha > 1$ and $\sigma_r = \sigma_z$.
Isotropy is the case when $\alpha = 1$. 
For anisotropic disk magnetic fields in M51, $\alpha$ has been measured to be $1.83$ by \citet{houde13} who measured the random field anisotropy in terms of the correlation scales in the two orthogonal directions ($x$ and $y$) and not in terms of the strength of the fluctuations in the two directions, as we use. 
For the halo anisotropic fields, $\alpha$ is expected to be less than the disk value as a result of weaker spiral density waves and differential rotation in the halo.
In our model, the disk and halo anisotropic factors are fixed to $2.0$ and $1.5$, respectively, and are reported in Table~\ref{modelparam}.  
Root mean square (rms) values are used for individual components of the turbulent magnetic field strengths in the disk or halo by normalizing the square isotropic $\sigma^2_{\text{I}}$ or anisotropic $\sigma^2_{\text{A}}$ field strength as $\sigma^2_r  = \sigma^2_{\text{I}} / 3$ for isotropy and $\sigma^2_r  = \sigma^2_{\text{A}} / (2 \, + \, \alpha) $ for anisotropy in Eq.~\eqref{sigma_eq}.

\subsection{Densities} \label{othermodelparams}

 The thermal electron density ($n_{\text{e}}$) is assumed to be a constant at each of the four radial rings and about an order of magnitude smaller in the halo than in the disk. 
Table~\ref{modelparam} displays these values along with the respective path lengths through the (flaring) disk and halo.
The cosmic ray density ($n_{\text{cr}}$) is assumed to be a global constant throughout the entire galaxy whose actual value is not significant as it cancels out upon computing $p$. 
Synchrotron emissivity is described as $\varepsilon = c B^2_{\perp}$ with constant $c = 0.1$.

\begin{table}
\caption{Model Standard Parameters. Thermal electron density ($n_{\text{e}}$) and path length ($L$) values are collected from \citet{berk} and \citet{fletcher11}. The parameter $\alpha$ is used to characterize anisotropic turbulent magnetic fields and is discussed in Section~\ref{turbumagfld}.}
\label{modelparam} 
\begin{center}
\begin{tabular}{ l  c  c  c} 
\hline\hline
\rule{0pt}{3ex} 
& $n_{\text{e}} \, [{\text{cm}}^{-3}]$ & $L$ [pc] & $\alpha$ \\ \hline 
Disk Ring 1,2 & 0.11 & 800 & 2.0 \\ 	
Disk Ring 3,4 & 0.06 & 1200 & 2.0 \\ 
Halo Ring 1,2 & 0.01 & 5000 & 1.5 \\ 
Halo Ring 3,4 & 0.006 & 3300 & 1.5 \\ 
\hline
\end{tabular}
\end{center}
\end{table}

\subsection{Depolarization} \label{depolarization}

We model the wavelength-dependent depolarization mechanisms of differential Faraday rotation (DFR) and internal Faraday dispersion (IFD) concomitantly to account for the presence of regular and turbulent magnetic fields in a given layer together with wavelength-independent depolarization. 
The combined wavelength-dependent and wavelength-independent depolarization for a two-layer system and three-layer system, with identical far and near sides of the halo, are given by \citep{shneiderI}
\begin{align} \label{directsum_2layer}
    \left(\frac{p}{p_{0}}\right)_{2layer} =  \Bigg\{  & W^2_d \, \left( \frac{I_d}{I} \right)^2 \left( \frac{1 - 2e^{-\Omega_d}\cos{C_d}+e^{-2\Omega_d}}{\Omega^2_d + C^2_d}\right) \nonumber \\
& +  W^2_h \, \left( \frac{I_h}{I} \right)^2 \left( \frac{1 - 2e^{-\Omega_h}\cos{C_h}+e^{-2 \Omega_h}}{\Omega^2_h + C^2_h}\right) 	\nonumber \\
& + W_d W_h \, \frac{I_d I_h}{I^2} \frac{2}{F^2 + G^2} \Bigg[ \left \lbrace F,G \right \rbrace \left( 2 \Delta \psi_{dh} + C_h \right) \nonumber \\
& + e^{-(\Omega_d \, + \, \Omega_h)} \left \lbrace F,G \right \rbrace \left( 2 \Delta \psi_{dh} + C_d \right) \nonumber \\
& - e^{- \Omega_d} \left \lbrace F,G \right \rbrace \left( 2 \Delta \psi_{dh} + C_d + C_h \right) \nonumber \\
& - e^{- \Omega_h}  \left \lbrace F,G \right \rbrace \left( 2 \Delta \psi_{dh} \right) \Bigg]  	\Bigg\}^{1/2},
\end{align}
and
 \begin{align} \label{directsum_3layer}
    & \left(\frac{p}{p_{0}}\right)_{3layer} = \Bigg(  2 \, W^2_h \, \left(\frac{I_h}{I} \right)^2 \Bigg\{ \left( \frac{1}{\Omega^2_h + C^2_h} \right) \, \times \nonumber \\
& \left(1 - 2e^{-\Omega_h}\cos{D}+e^{-2\Omega_h}\right) \Big[ 1 + \cos \left( C_d + C_h \right) \Big]  \Bigg\} \nonumber \\
& +  W^2_d \, \left(\frac{I_d}{I} \right)^2 \left( \frac{1 - 2e^{-\Omega_d}\cos{C}+e^{-2\Omega_d}}{\Omega^2_d + C^2_d}\right)	\nonumber \\
& + W_d W_h \, \frac{I_d I_h}{I^2} \frac{2}{F^2 + G^2} \Bigg\{ \left \lbrace F,-G \right \rbrace \left( -2 \Delta \psi_{dh} + C_d \right) \nonumber \\
&+ \left \lbrace F,G \right \rbrace \left( 2 \Delta \psi_{dh} + C_h \right) \nonumber \\
& + e^{-(\Omega_d \,+ \, \Omega_h)} \Big[ \left \lbrace F,G \right \rbrace \left( 2 \Delta \psi_{dh} + C_d \right) + \left \lbrace F,-G \right \rbrace \left( -2 \Delta \psi_{dh} + C_h \right) 	 \Big] \nonumber \\
& - e^{-\Omega_d} \Big[ \left \lbrace F,G \right \rbrace \left( 2 \Delta \psi_{dh} + C_d + C_h \right) + \left \lbrace F,-G \right \rbrace \left( -2 \Delta \psi_{dh} \right) \Big] \nonumber \\
& - e^{-\Omega_h} \Big[ \left \lbrace F,-G \right \rbrace \left( -2 \Delta \psi_{dh} + C_d + C_h \right) + \left \lbrace F,G \right \rbrace \left( 2 \Delta \psi_{dh} \right) \Big] \Bigg\} \Bigg)^{1/2},
 \end{align}
where $p_0$ is the intrinsic degree of linear polarization of synchrotron radiation, $\{d,h\}$ denote the disk and halo, $\Omega_d = 2 \sigma^2_{RM_d} \lambda^4 $, $\Omega_h = 2 \sigma^2_{RM_h} \lambda^4$, $C_d = 2 R_d \lambda^2$, $C_h = 2 R_h \lambda^2$, $F = \Omega_d \Omega_h \,+ \,C_d C_h$, $G = \Omega_h C_d \,- \, \Omega_d C_h$.  

In Eqs.~\eqref{directsum_2layer} and~\eqref{directsum_3layer}, the per-layer total synchrotron emission $I_i$, the total Faraday depth $R_i$, the dispersion of the intrinsic RM within the volume of the telescope beam $\sigma_{RM_i}$, along with the wavelength-independent depolarizing terms $W_i$ are respectively given as
\begin{align} 
   I_i & = \varepsilon_i \, L_i, \nonumber \\
   R_i & = 0.81 \, {n_{\text{e}i}} \, \overline B_{\parallel i} \, L_i,	\nonumber \\
   \sigma_{RM_i} &= 0.81 \, \left\langle n_{\text{e}i} \right\rangle \, b_{\parallel i} \left(L_i \, d_i \right)^{1/2},   \label{sigma_RM} \\
   W_i & = \left\{ \frac{\left \lbrack \left( \overline B^2_x - \overline B^2_y + \sigma^2_x - \sigma^2_y \right)^{2} + 4 \overline B^2_x  \overline B^2_y \right \rbrack^{1/2}}{\overline{B^2_\perp}} \right\}_i  \label{gen_form} , 
\end{align}
where $\varepsilon_i$ is the synchrotron emissivity, $I_i$ is the synchrotron intensity, $L_i$ is the path length (pc), along with  
$\overline{B}^2_\perp = \overline B^2_x \,+ \, \overline B^2_y$ and $\overline{B^2_\perp} = \overline B^2_\perp \, + \, \sigma^2_x \, + \, \sigma^2_y $.
The form of $W_i$ in Eq.~\eqref{gen_form} implicitly assumes that emissivity scales with $\varepsilon \propto B^2_{\perp}$ corresponding to a synchrotron spectral index of -1. 
Isotropic expressions for the intrinsic polarization angle and for wavelength-independent depolarization are obtained by setting $\sigma_x = \sigma_y$.
The operation $ \left \lbrace F, G \right \rbrace \left(a \right)$ is defined as $ \left \lbrace F, G \right \rbrace \left(a \right) = F\cos \left(a \right) \,-\, G\sin \left(a \right)$.
$\Delta \psi_{dh} = \left\langle\psi_{0d}\right\rangle \, - \, \left\langle\psi_{0h}\right\rangle$ is the difference in the projected intrinsic polarization angles of the disk and halo with the respective angles given by \citep{sokol,sokol_erratum} as
\begin{equation} \label{psi_gen}
 \left \langle \psi_{0i} \right \rangle = \tfrac{\pi}{2} \, - \, \arctan \left[\cos(l) \tan(\phi) \right] \, + \, \tfrac{1}{2} \arctan \left( \frac{2 \overline B_x \overline B_y}{\overline B^2_x - \overline B^2_y + \sigma^2_x - \sigma^2_y} \right)_i.
 \end{equation}
Expectation values denoted by $\langle \ldots \rangle$ arise whenever turbulent magnetic fields are present. 
Only the last term of Eq.~\eqref{psi_gen} remains upon taking the difference. 

In our use of Eq.~\eqref{sigma_RM} to describe both isotropic and anisotropic random fields we implicitly treat $\sigma_{RM}$ as a global constant, independent of the observer's viewing angle as for a purely isotropic random field. 
Moreover, the diameter of a turbulent cell $d_i$ in the disk or halo, as it appears in Eq.~\eqref{sigma_RM}, is approximately given by \citep{fletcher11}
\begin{equation} \label{diam_turb_cell}
d_i \simeq \left[ \frac{D \,\sigma_{RM,D}}{ 0.81 \, \left \langle n_{\text{e}i} \right \rangle  \, b_{\parallel i} \, (L_i)^{1/2}} \right]^{2/3}, 
\end{equation}
with $\sigma_{RM,D}$ denoting the RM dispersion observed within a telescope beam of a linear diameter $D = 600$ pc.  
$\sigma_{RM,D}$ has been fixed to the observed value of $15 \, \text{rad m}^{-2}$ \citep{fletcher11}.

\section{Procedure} \label{procedure}

 We use various magnetic field configurations of isotropic turbulent and/or anisotropic turbulent fields in the disk and halo with the requirement that there be at least a turbulent magnetic field in the disk following \citet{fletcher11} observations. 
We also model wavelength-independent depolarization directly via $W_i$ in Eq.~\eqref{gen_form} instead of approximating it with the value of $p$ at the shortest wavelength.
Consequently, these turbulent configurations, given in Table~\ref{model_types}, span 12 of the 17 model types listed in \citet[upper panel of Table 2]{shneiderI} and are illustrated in their Figs. 2 and 3 for an example bin with a particular choice of magnetic field strengths.
These configurations may also be viewed in terms of two distinct groups characterized by the presence or absence of a turbulent magnetic field in the halo. 

\begin{table}
\caption{Model settings for a two- or three-layer system based on regular and turbulent magnetic field configurations in the disk and halo.}	
\label{model_types}
\begin{center}
\begin{tabular}{l c c c c c c } 
\hline 
\hline 
\rule{0pt}{2.5ex}
& \multicolumn{3}{ c }{Disk} &  \multicolumn{3}{ c }{Halo} \\  
\hline
\rule{0pt}{2.5ex}
& Reg. & Iso. & Aniso. & Reg. & Iso. & Aniso. \\ 
\hline 
DIH					&$\checkmark$&$\checkmark$&		&$\checkmark$&	&	\\
DAH					&$\checkmark$&	&$\checkmark$&$\checkmark$&		& 	\\	
DAIH					&$\checkmark$&$\checkmark$&$\checkmark$&$\checkmark$&	& 	\\
DIHI					&$\checkmark$&$\checkmark$&		&$\checkmark$&$\checkmark$&	\\
DIHA					&$\checkmark$&$\checkmark$& 	&$\checkmark$&	&$\checkmark$	\\	
DAHI					&$\checkmark$&	&$\checkmark$&$\checkmark$&$\checkmark$&	\\
DAHA					&$\checkmark$&	&$\checkmark$&$\checkmark$&	&$\checkmark$	\\	
DIHAI					&$\checkmark$&$\checkmark$& 	&$\checkmark$&$\checkmark$&$\checkmark$	\\
DAHAI					&$\checkmark$&	&$\checkmark$&$\checkmark$&$\checkmark$&$\checkmark$	\\
DAIHI					&$\checkmark$&$\checkmark$&$\checkmark$&$\checkmark$&$\checkmark$&	\\
DAIHA					&$\checkmark$&$\checkmark$&$\checkmark$&$\checkmark$&	&$\checkmark$	\\
DAIHAI					&$\checkmark$&$\checkmark$&$\checkmark$&$\checkmark$&$\checkmark$&$\checkmark$	\\
\hline	
\end{tabular} 
\tablefoot{ The three column headings below the principle headings of the `Disk' and `Halo' denote the regular, isotropic turbulent, and anisotropic turbulent magnetic fields. 
The rows contain a listing of all model types simulated with the following nomenclature: `D' and `H' denote disk and halo magnetic fields, respectively, `I' and `A' are the isotropic and anisotropic turbulent magnetic fields.
}
\end{center}
\end{table}

The isotropic and anisotropic turbulent magnetic field strengths in the disk and halo are each sampled from $[0,2,5,8,10,15,20,25,30] \, \mu$G
in line with M51 observations of having a $10 \, \mu$G isotropic and a $10 \, \mu$G anisotropic turbulent field in the disk \citep{houde13}. 
We assume that the total turbulent field strength in the halo is less than or equal to that in the disk. 
For each of these turbulent magnetic field configurations, we allow the regular magnetic fields in the disk and halo to separately vary in the ranges of $0 - 50 \, \mbox{$\mu G$}$ in steps of $0.1 \, \mbox{$\mu G$}$.

We apply a reduced chi-square statistic to discern a best-fit magnetic field configuration for each of the four radial rings, independently, at the three observing wavelengths $\lambda \lambda \lambda \, 3.5, 6.2, 20.5$ cm.
The reduced chi-square statistic is given by
  \begin{equation*} \label{chisq_red}
  \chi^{2}_{red} = \frac{\chi^{2}}{N} \, = \, \frac{1}{N} \sum_{bins \, \in \, ring} \frac {\left( p_{obs} - p_{mod} \right){}^2}{\sigma^2},
  \end{equation*} 
where $p_{obs}$ and $p_{mod}$ are the observed and modeled $p$ values given in Eqs.~\eqref{directsum_2layer} and~\eqref{directsum_3layer}, $\sigma$ is the standard deviation of the measured $p$ values per bin in a given ring, and the sum is taken over all bins comprising a given ring. $N$ is the number of degrees of freedom given by
$\left(\textit{\# observing wavelengths}\right) \times \left(\textit{\# bins in a ring}\right) - \left( \textit{\# independent parameters}\right)$,
with the  number of independent parameters being the variable disk and halo regular magnetic field strengths and, hence, always two, for a fixed input of turbulent magnetic fields describing a particular configuration. 

For each turbulent magnetic field configuration sampled, the best-fit combination of total disk and halo regular magnetic field strengths corresponding to the lowest $\chi^{2}_{red}$ value are found and a range of $\chi^{2}_{red}$ contours are plotted in order to examine the $\chi^{2}_{red}$ landscape. Repeating this procedure allows for a global minimum $\chi^{2}_{red}$ value to be obtained for each of the rings. 

\begin{figure*}
\centering
\includegraphics[width=.45\hsize]{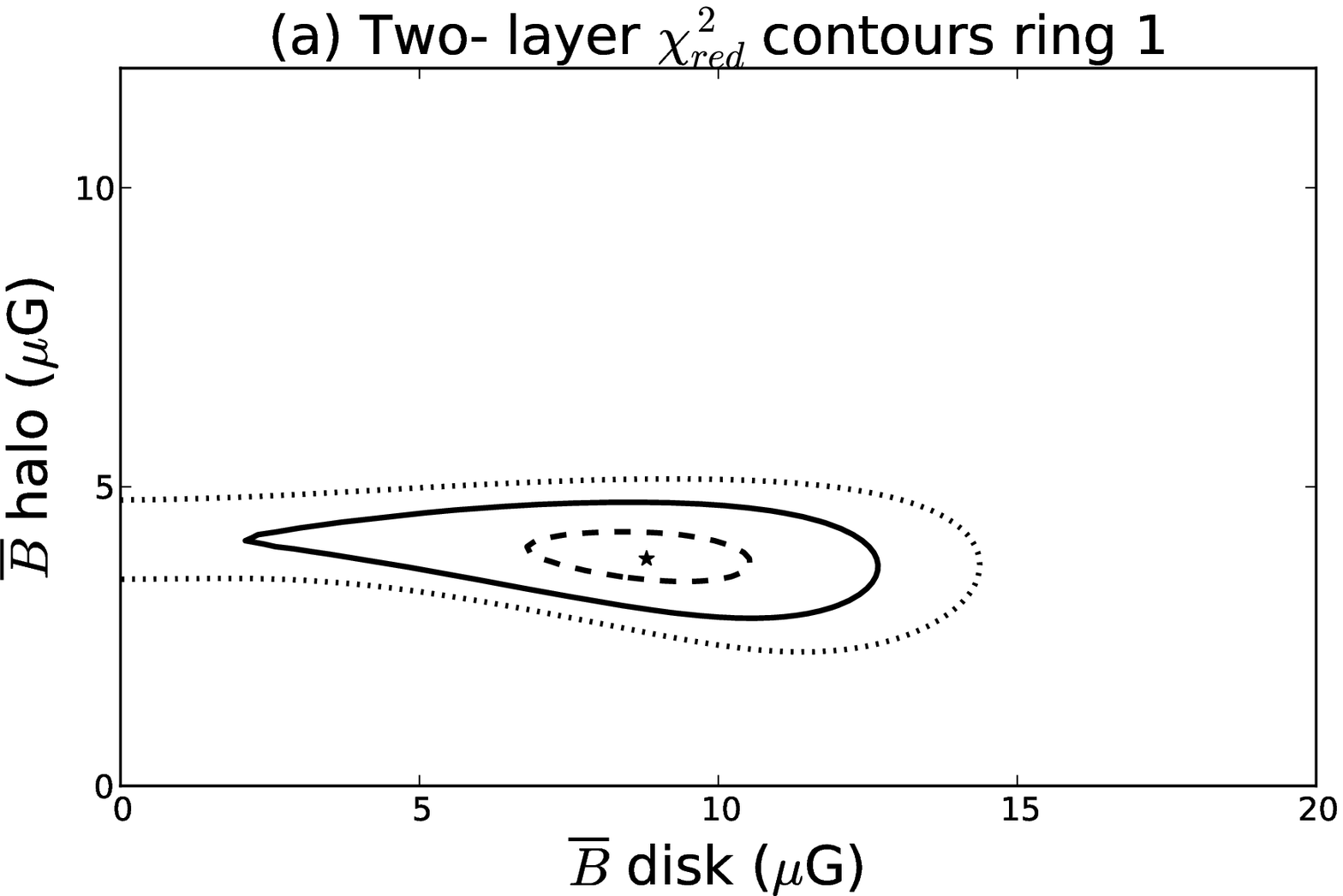}~~~~~~
\includegraphics[width=.45\hsize]{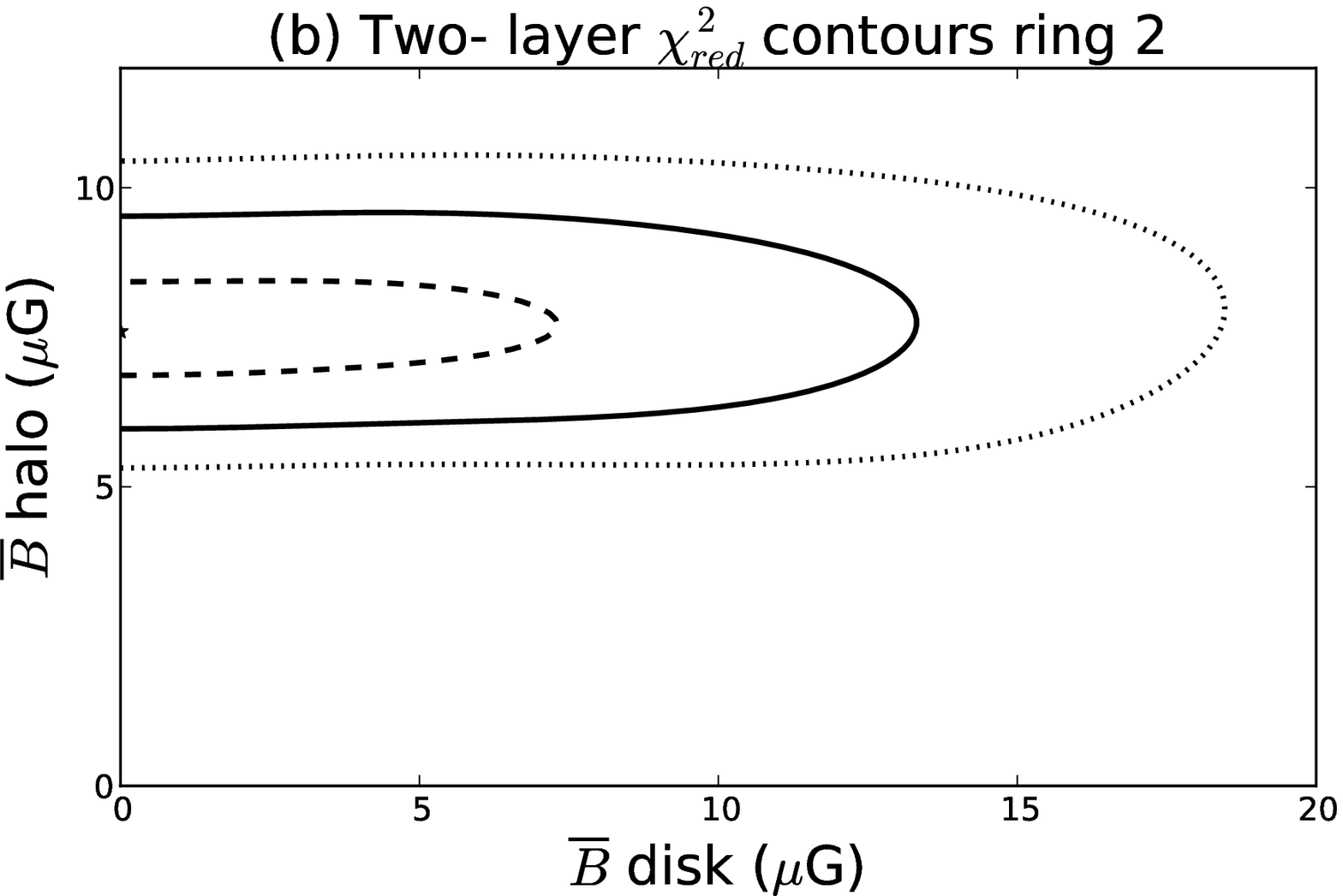} \vspace{1.0cm} \\  
\includegraphics[width=.45\hsize]{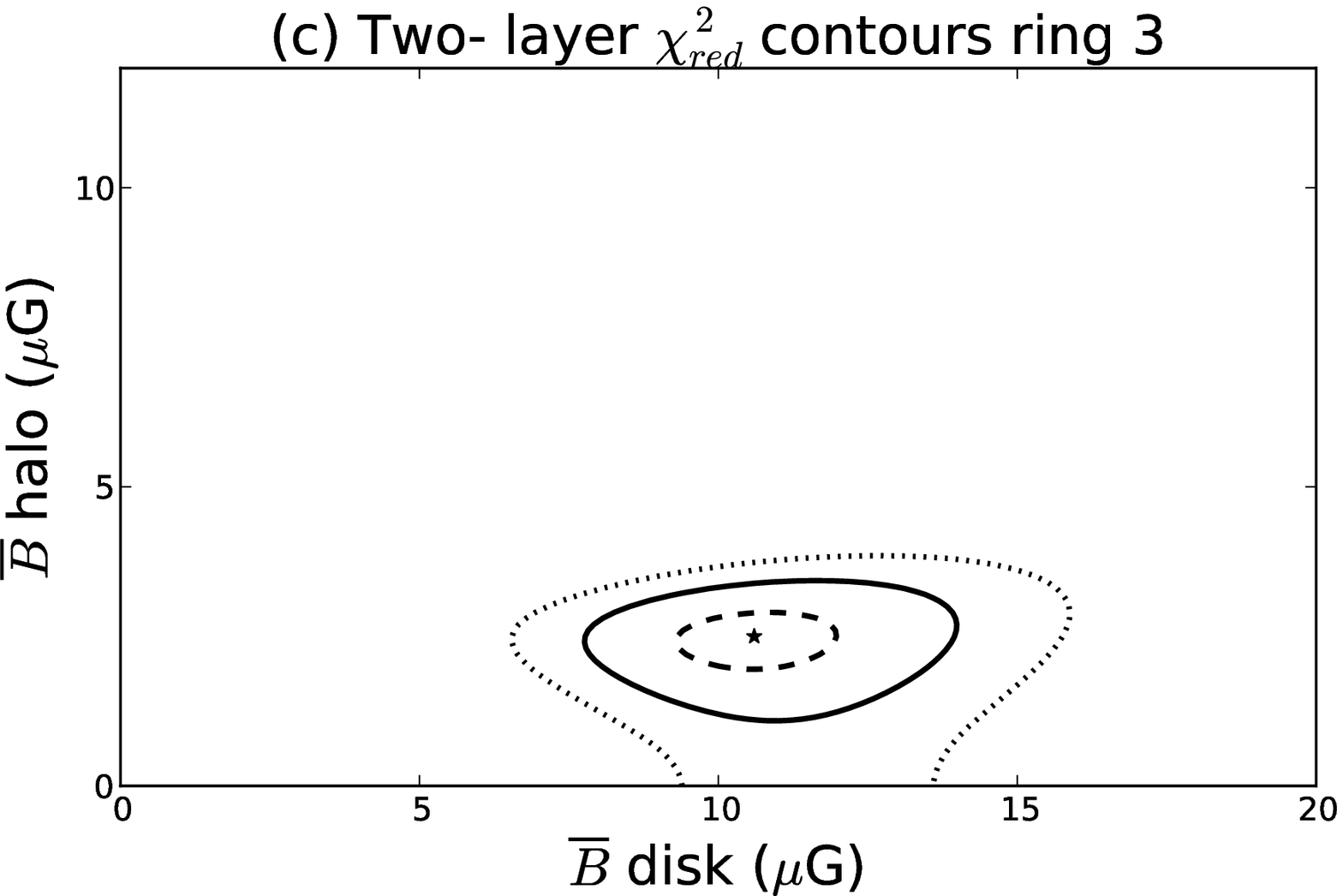}~~~~~~
\includegraphics[width=.45\hsize]{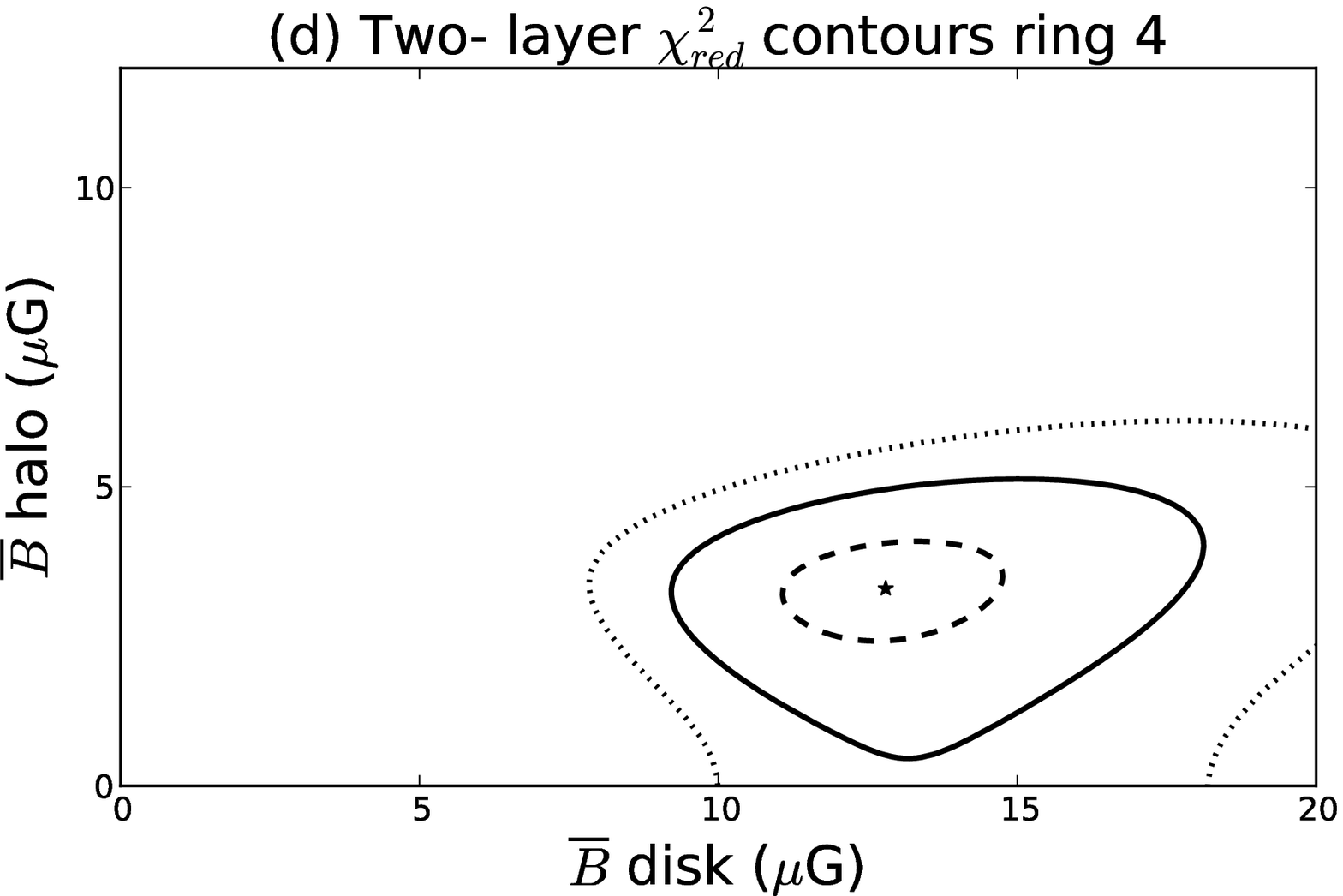}	
\caption{(a)-(d) Contours of equal reduced chi-square values for regular magnetic field strengths $\overline{B}$ in disk and halo in a two-layer model for each of the four rings. 
The best-fit DAIHI model, denoted by $\bigstar$, is composed of regular, isotropic turbulent and anisotropic turbulent disk and halo magnetic fields with respective minimum reduced chi-square ($\chi^{2}_{min}$) values and field strengths presented in Table~\ref{two_layer_strengths_cell_sizes}. 
The dashed, solid, and dotted contours represent $10, 50, \, \text{and} \, 100$ percent increases in the $\chi^{2}_{min}$ value, respectively.}
\label{Chi_ring1_2_3_4} 
\end{figure*}

$\chi^{2}_{red}$ values larger than one are accepted in order to establish a trend in turbulent magnetic field configurations and strengths. 
To test whether the admission of these higher $\chi^{2}_{red}$ values yield regular disk and halo magnetic field configurations that are statistically consistent for each ring, we use a generalization error approach (bootstrap technique) which is independent of the $\chi^{2}_{red}$ statistic. This approach stipulates to approximately retain $70\%$ of the data while discarding around $30\%$ of the data at random, for each independent trial run, and to check the resulting fits again. In this way, the stability of the lowest $\chi^{2}_{red}$ contours for a particular configuration is tested.  
Following 50 such independent trial runs for each of the global $\chi^{2}_{red}$ minimum found per ring reveals that all such lowest $\chi^{2}_{red}$ contours are \emph{stable} for both a two-layer model and (quasi) stable for a three-layer model.

We examine a smaller subset of the turbulent field configurations for a three-layer model making sure to examine configurations that are both good and poor fits for the corresponding two-layer system.

\section{Results} \label{results}

\subsection{Two-layer model} 

\begin{table} 
\caption{Two-layer best-fit DAIHI model magnetic field strengths. Values in parenthesis correspond to the alternative best-fit model adopted for ring 2.}	
\label{two_layer_strengths_cell_sizes} 
\begin{center}
\tabcolsep=0.11cm
\begin{tabular}{ l  c  c  c  c }
\hline\hline
\rule{0pt}{2.5ex} 
& Ring 1 & Ring 2 & Ring 3 & Ring 4 \\ 
\hline
\rule{0pt}{2.5ex}
Disk &  &  &  & \\
\hline
\rule{0pt}{2.5ex}
Iso.[$\mu$G] 	& $10$ 	& $10$ 	& $10$ 	& $10$  \\
Aniso.[$\mu$G]	& $5$	& $10$	& $10$	& $10$	\\
Reg.[$\mu$G] 	& ${8.8}^{+4}_{-7}$	& ${0.0}^{+13}_{-0}$ $({12.4}^{+5}_{-4})$		& $10.6 \pm 3$	&	${12.8}^{+5}_{-4}$ \\
d[pc] 		& $47$ 	& $40$ 	& $52$ 	& $52$ \\
\hline
\rule{0pt}{2.5ex}
Halo &  &  &  & \\
\hline	
\rule{0pt}{2.5ex}
Iso.[$\mu$G] 	& $5$	& $10$ $(2)$	& $2$	& $2$	\\
Aniso.[$\mu$G]	& $0$	& $0$	& $0$	& $0$	\\
Reg.[$\mu$G] 	& $3.8 \pm 1$	& $7.6 \pm 2$ $(1.5 \pm 1.5)$	& $2.5 \pm 1$	&	${3.3}^{+2}_{-3}$ \\
d[pc] & $215$ & $135$ $(395)$ & $638$ & $638$ \\
\hline
\rule{0pt}{2.5ex}
$\chi^{2}_{min}$ & $1.2$ & $2.4$ $(3.1)$ & $2.1$ & $3.0$ \\ 
\hline
\end{tabular} 
\end{center}
\end{table}

\begin{figure}
\centering 
\includegraphics[width=.95\hsize]{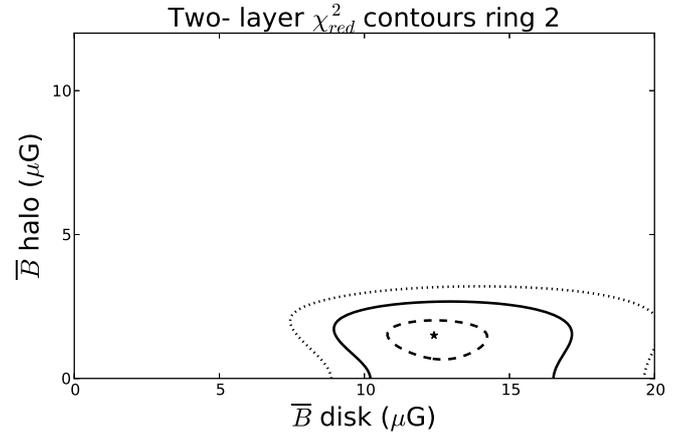} 
\caption{Contours of constant $\chi_{\text{min}}^2$ for values of regular field in the disk and halo for ring 2 with a deviating value for the isotropic turbulent field corresponding to the alternative best-fit model adopted, see text. 
Symbols are the same as used in Fig.~\ref{Chi_ring1_2_3_4}.} 
\label{chi_ring2_revised}
\end{figure}

The turbulent magnetic field strengths which correspond to the best-fit two-layer model per ring are presented in Table~\ref{two_layer_strengths_cell_sizes} together with the best-fit regular disk and halo field strengths attained from the reduced chi-squared analysis.
Errors reported for these respective regular field strengths are based on the solid contour in Fig.~\ref{Chi_ring1_2_3_4} which represents a $50\%$ increase in the $\chi^{2}_{min}$ value. 
$\chi^{2}_{min}$ is the minimum $\chi^{2}_{red}$ value corresponding to the best-fit disk and halo magnetic field configuration composed of regular, isotropic turbulent, and anisotropic turbulent magnetic fields.

\begin{figure}	
\centering		
\includegraphics[width=\hsize]{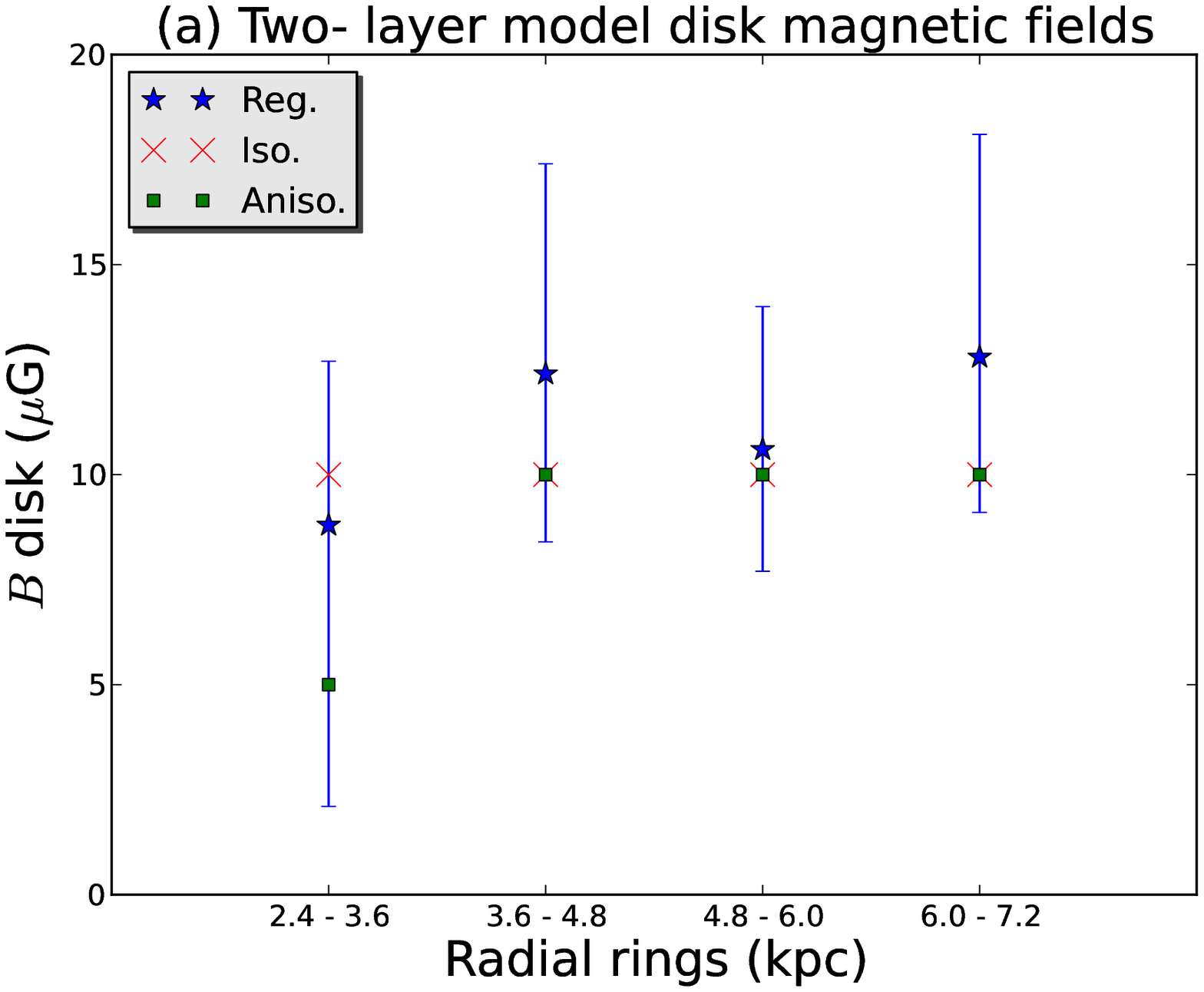}
\includegraphics[width=\hsize]{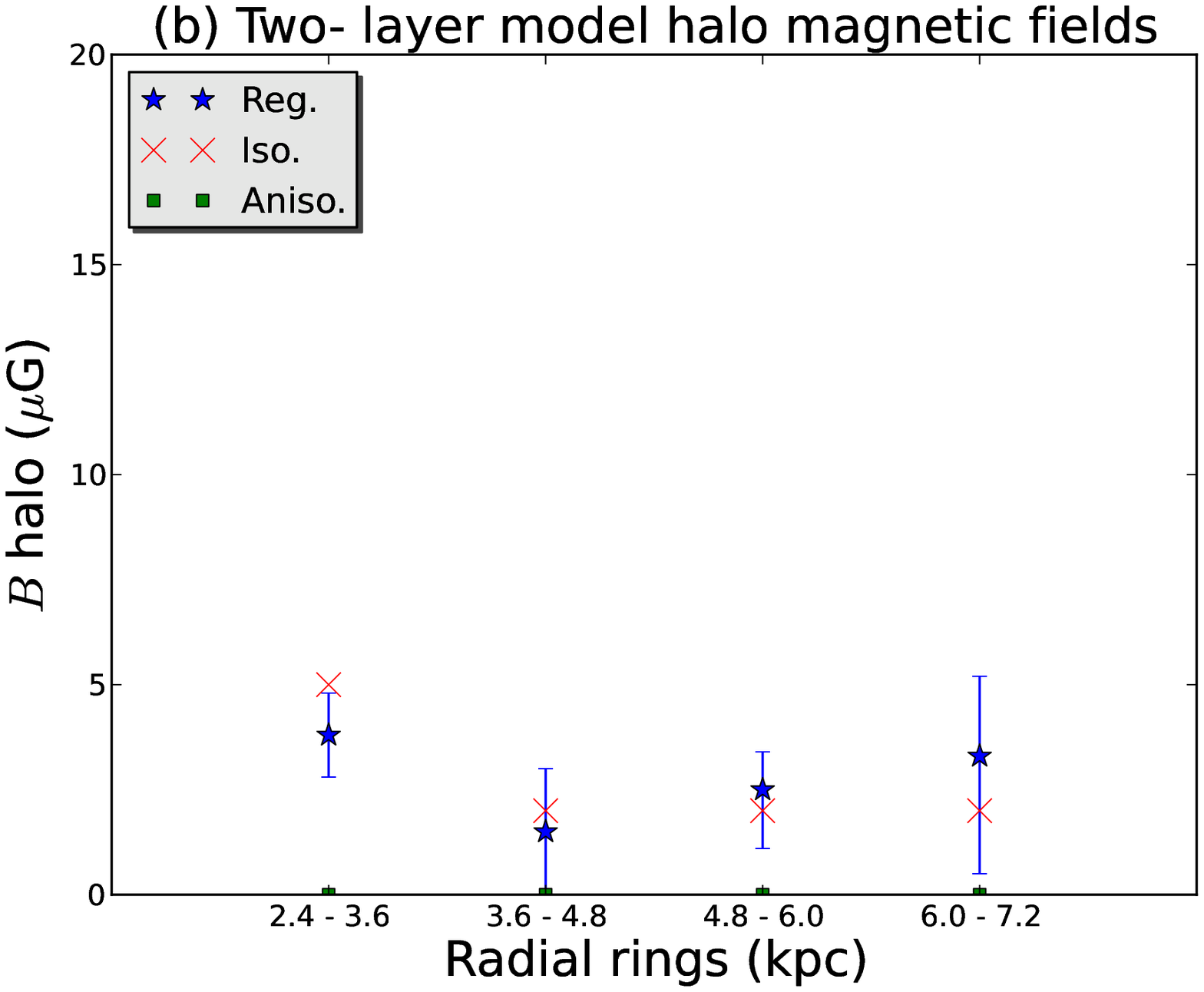}
\caption{Predicted magnetic field strengths ($\mu$G) with radial distance (kpc) from M51. The best-fit two-layer model configuration consisting of an isotropic turbulent (`Iso.'), anisotropic turbulent (`Aniso.'), and regular (`Reg.') magnetic field strengths in the disk (a) and halo (b) is shown per ring.}
\label{schematic_disk_halo_flds}
\end{figure}

Figure~\ref{Chi_ring1_2_3_4} and Table~\ref{two_layer_strengths_cell_sizes} clearly indicate that the best-fit magnetic field values in ring 2 deviate from the trend in the other three rings, especially $B_{\text{reg}}$ in the disk and $B_{\text{iso}}$ in the halo. To test how significant this deviation from the other rings is, we calculated a best-fit model with magnetic field values consistent with the other rings and checked how much the $\chi_{\text{red}}^2$ increased. Inserting $B_{\text{iso}} = 2~\mu$G in the halo for ring 2, results in a minimum $\chi_{\text{red}}^2 = 3.1$ for best-fit regular field values of $12.4~\mu$G and $1.5~\mu$G in the disk and halo, respectively (see Fig.~\ref{chi_ring2_revised}). Considering the uncertainties in the model, an increase in $\chi_{\text{min}}^2$ from 2.4 to 3.1 is not believed to be a significant difference in ring 2. We conclude that these field values are equally plausible and choose to adopt them as the best-fit model, making all magnetic field values in all rings roughly consistent.   
Fig.~\ref{schematic_disk_halo_flds} illustrates these regular and turbulent magnetic field values for the two-layer best-fit models. 

Global conclusions to be drawn from these magnetic field values are:
\begin{itemize}
\item The total magnetic field strength in the disk is about $B_{tot,disk} \approx 18~\mu$G, while the total magnetic field strength in the halo is about  $B_{tot,halo} \approx 4-6~\mu$G;
\item Both regular and turbulent magnetic field strengths in the disk are a few times higher than those in the halo;
\item There is a significant anisotropic turbulent field component in the disk, but not in the halo;
\item Within the errors, none of the magnetic field strengths shows a clear trend as a function of galactocentric radius. A possible exception here is a slightly stronger (isotropic) random magnetic field strength in the inner halo.
\end{itemize}

\begin{figure*}
\centering	
\includegraphics[width=.30\hsize]{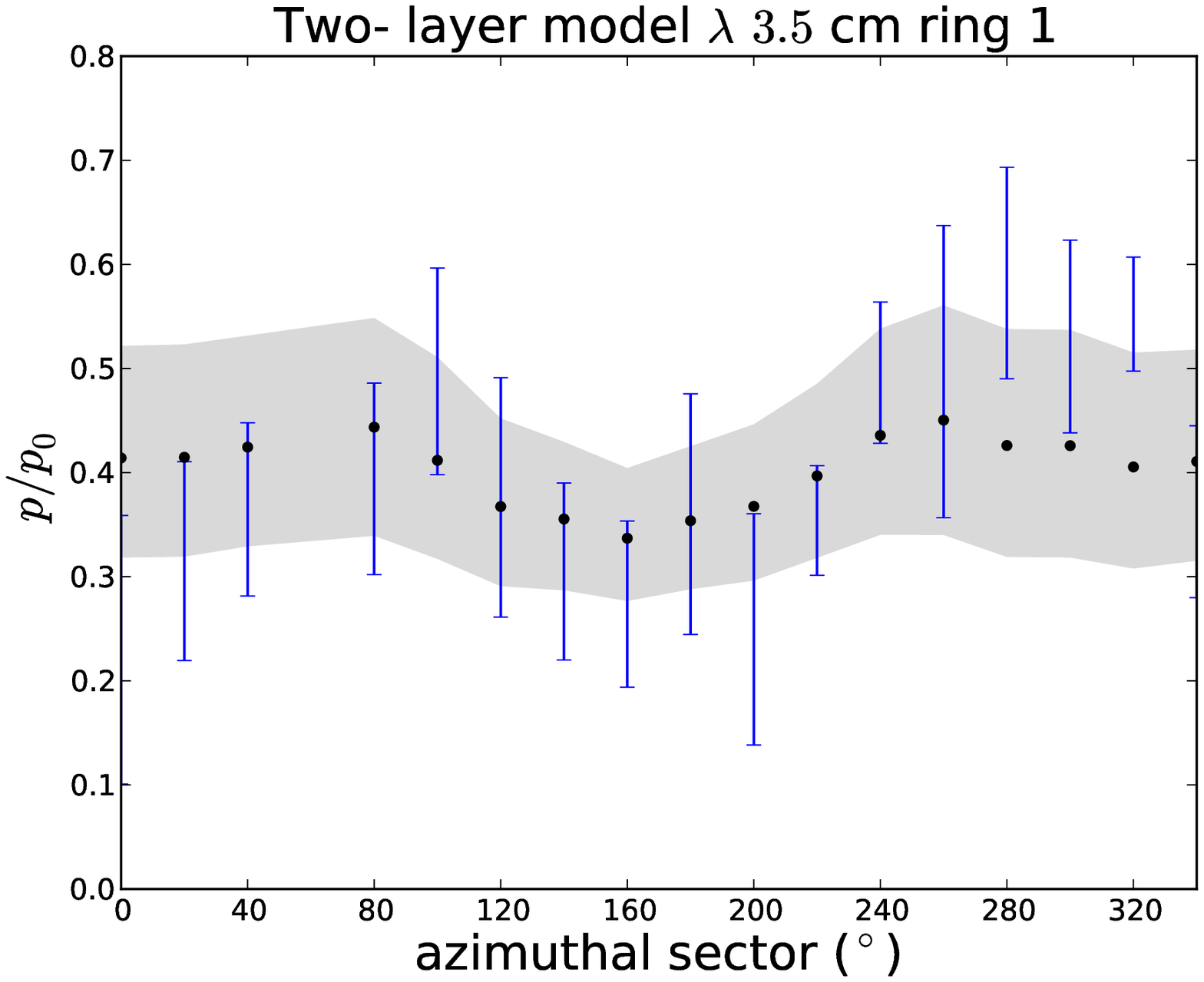}~
\includegraphics[width=.30\hsize]{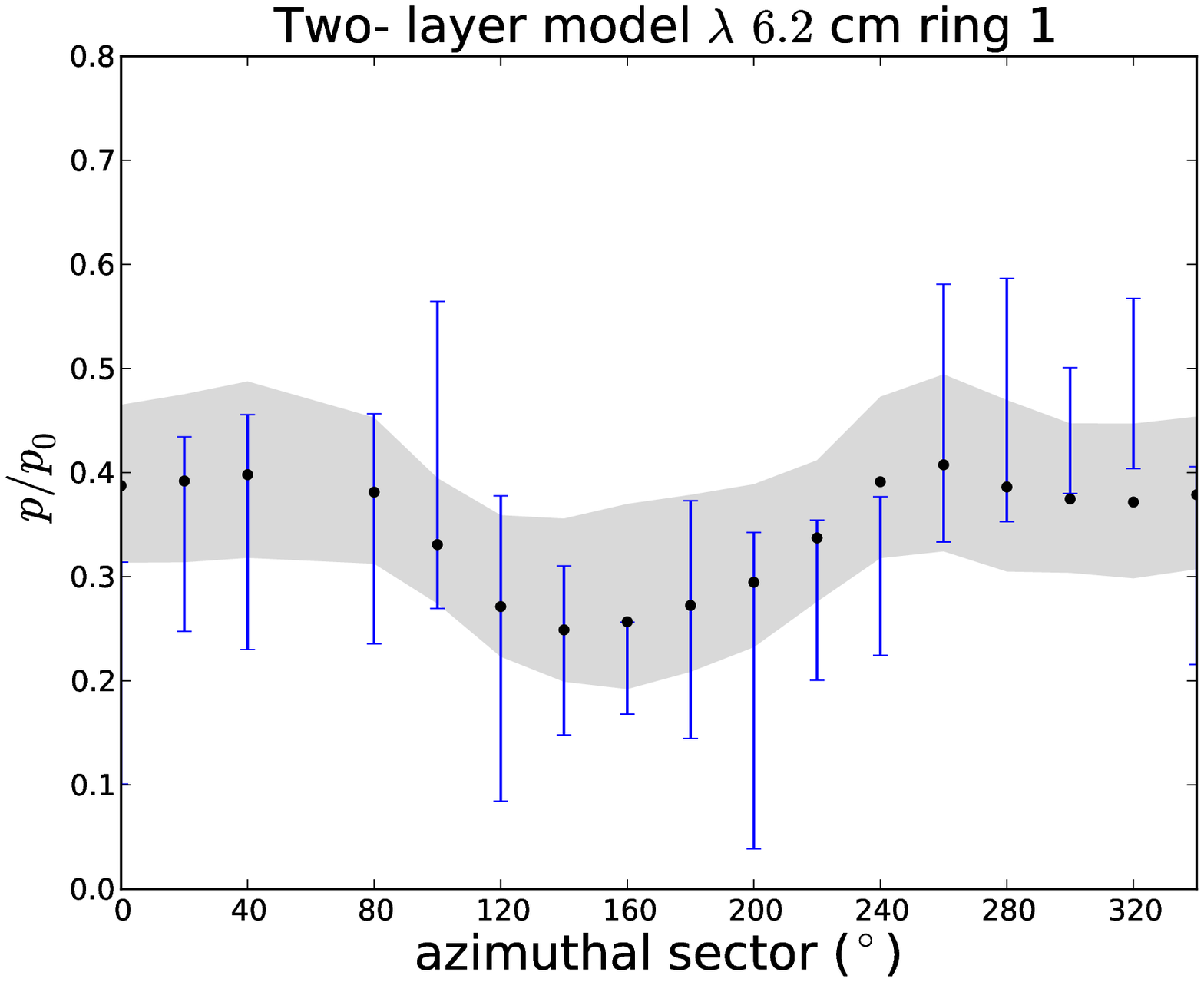}~
\includegraphics[width=.30\hsize]{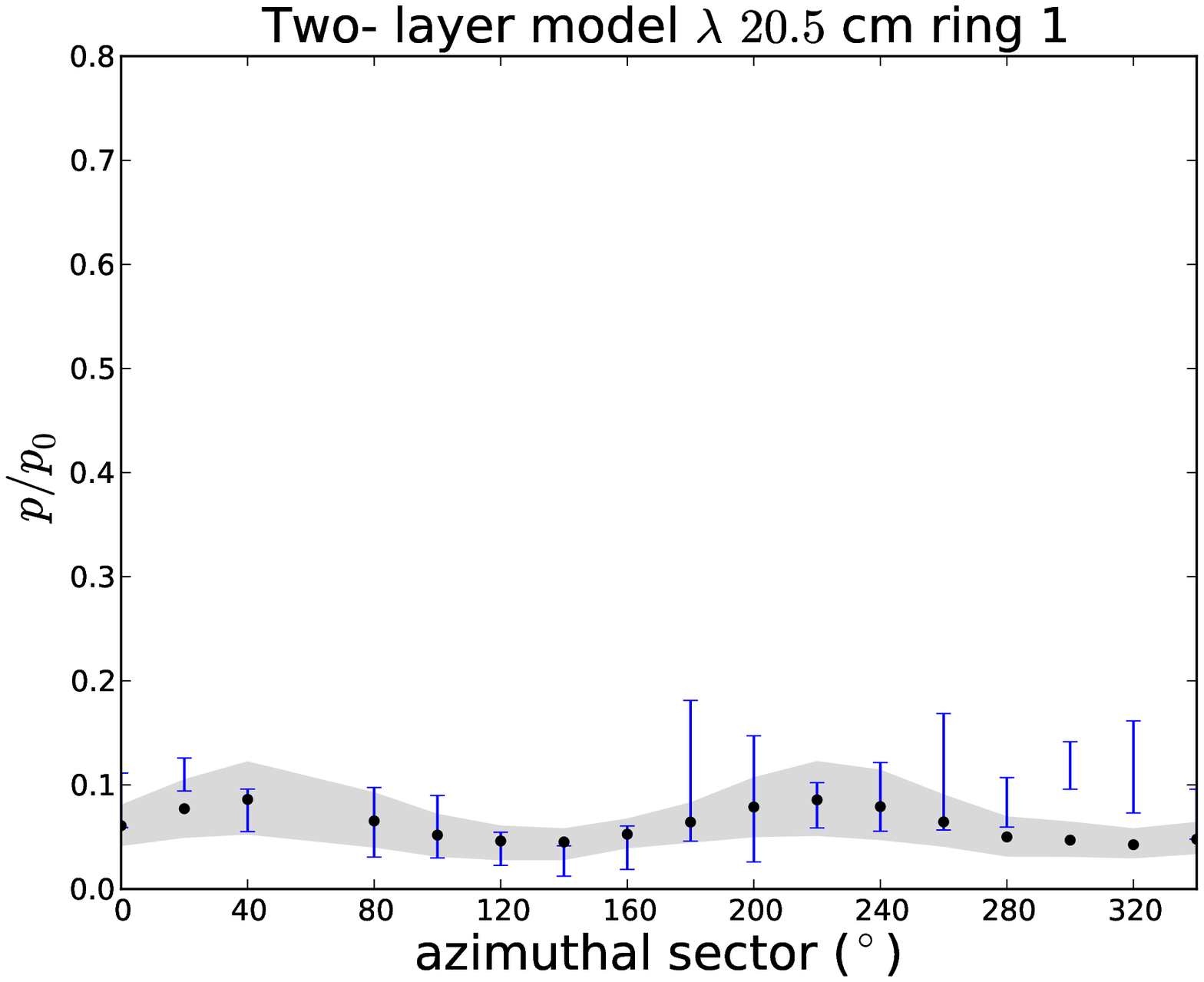} \\
\includegraphics[width=.30\hsize]{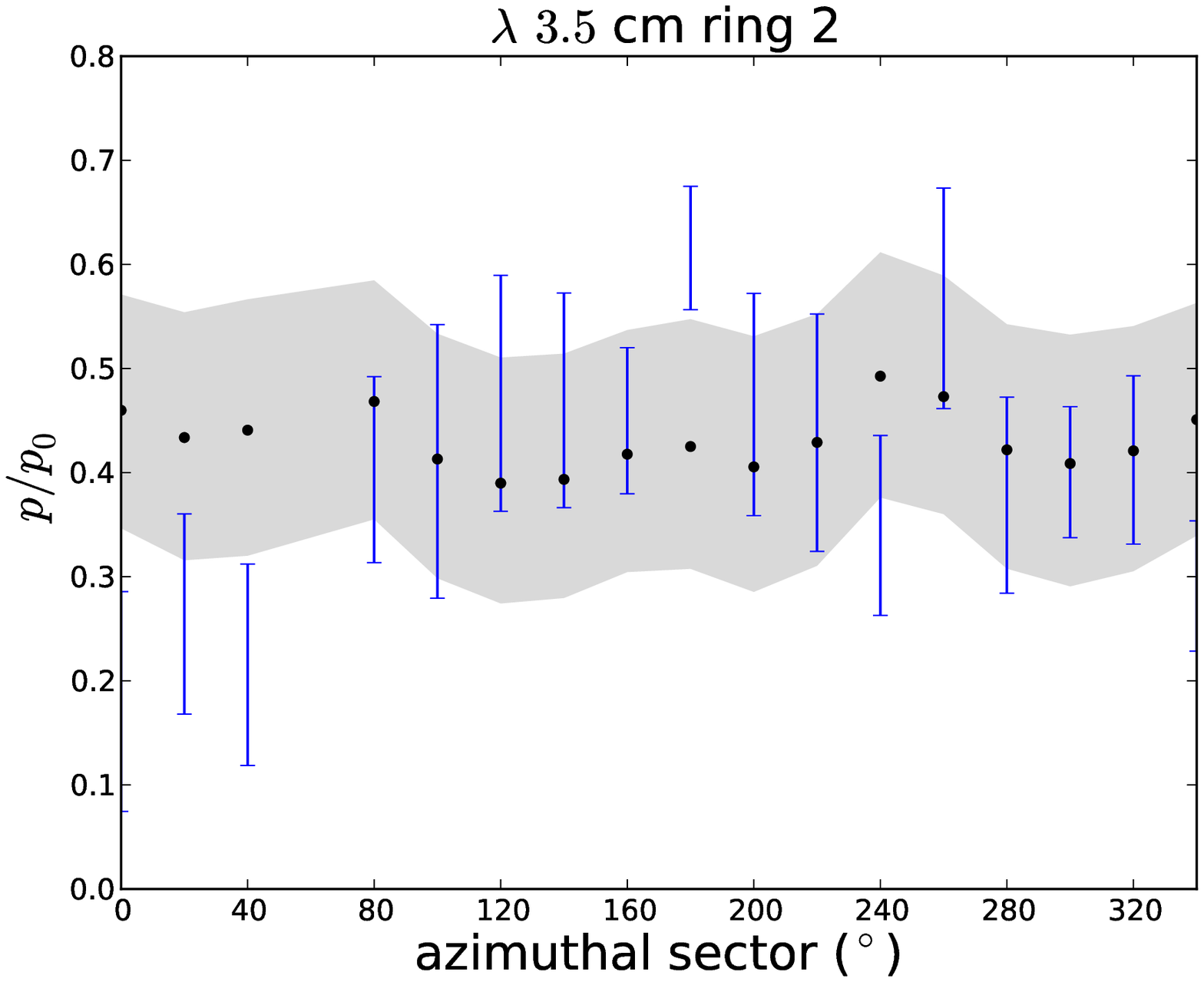}~ 
\includegraphics[width=.30\hsize]{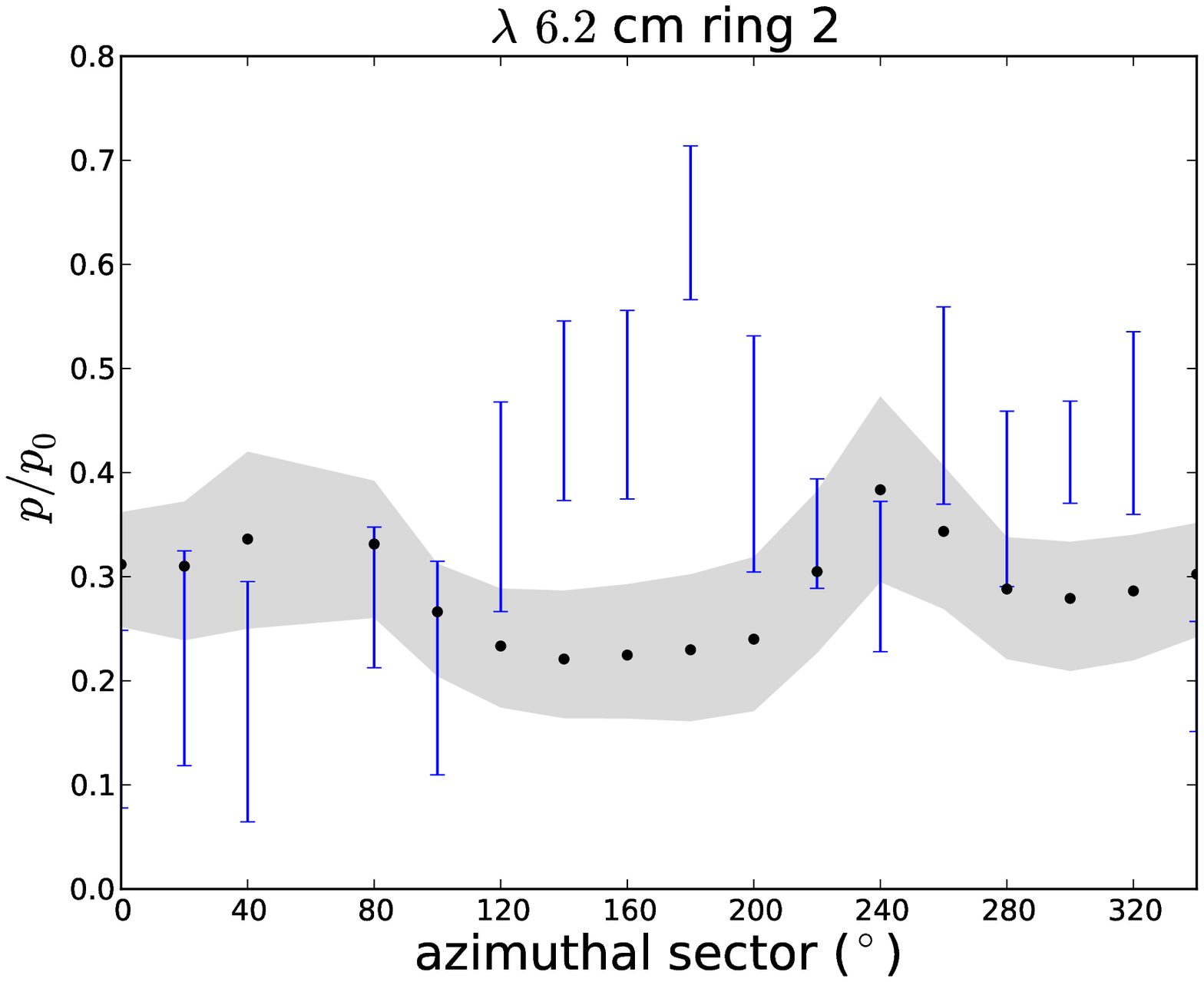}~ 
\includegraphics[width=.30\hsize]{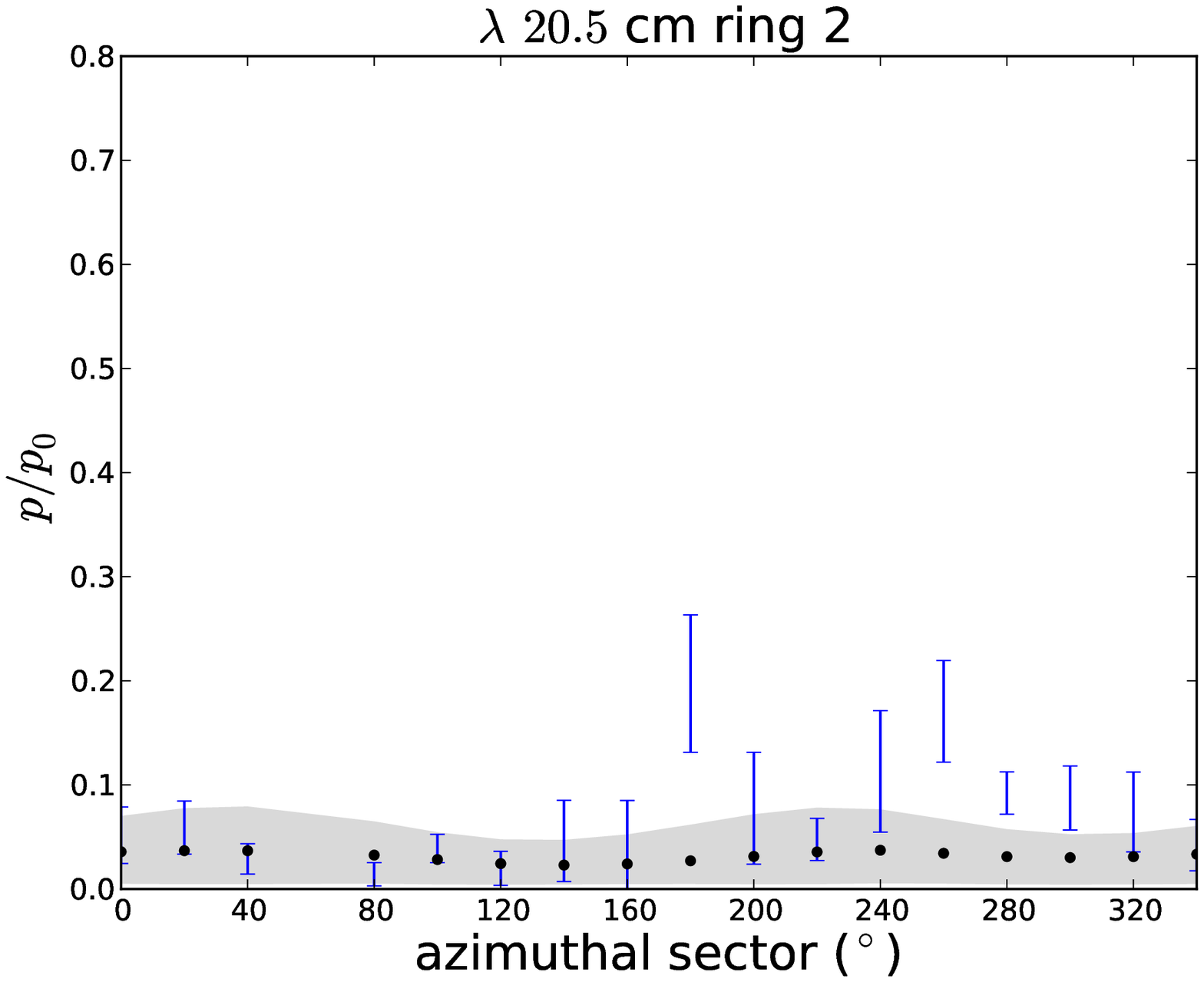} \\ 
\includegraphics[width=.30\hsize]{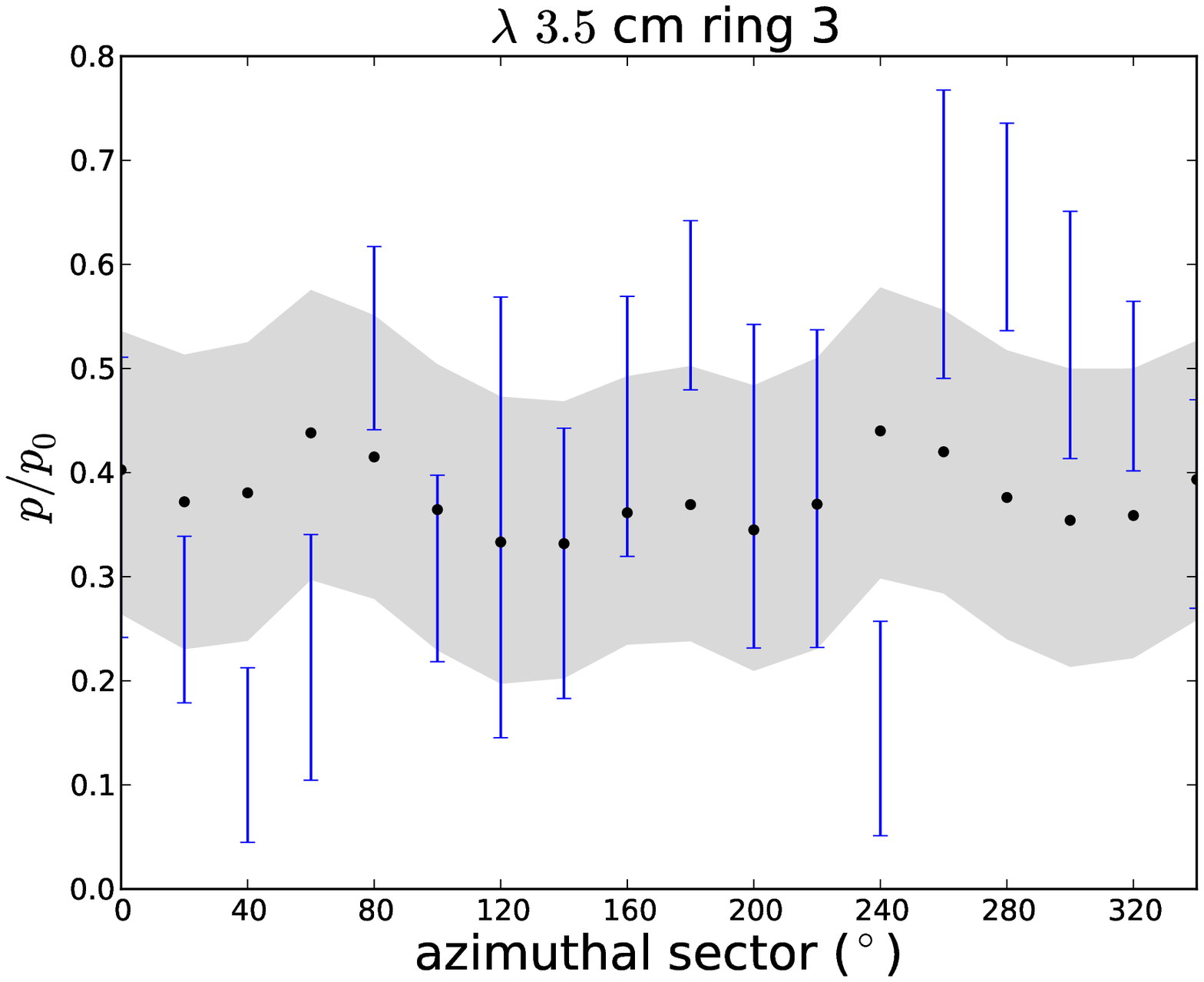}~
\includegraphics[width=.30\hsize]{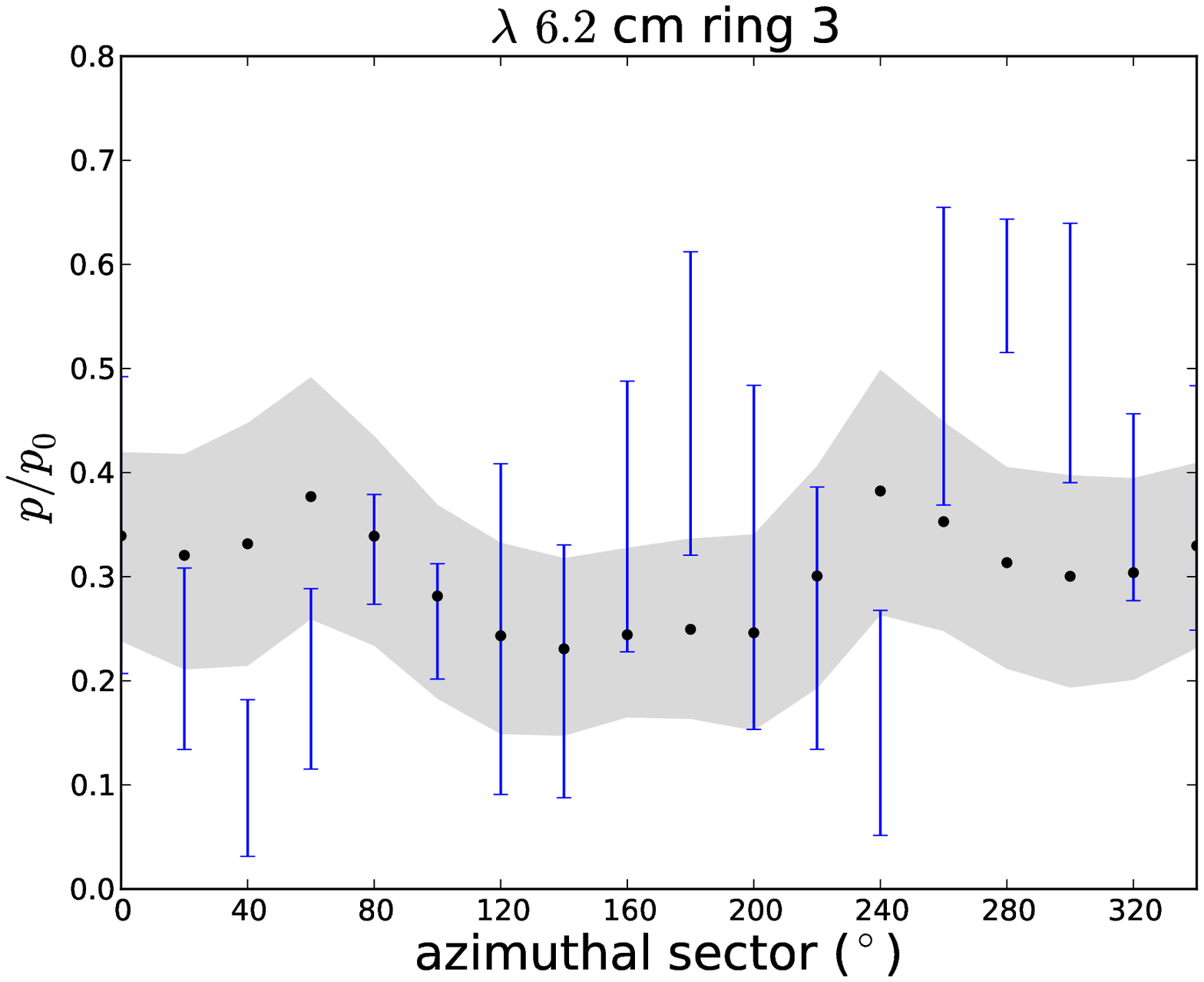}~
\includegraphics[width=.30\hsize]{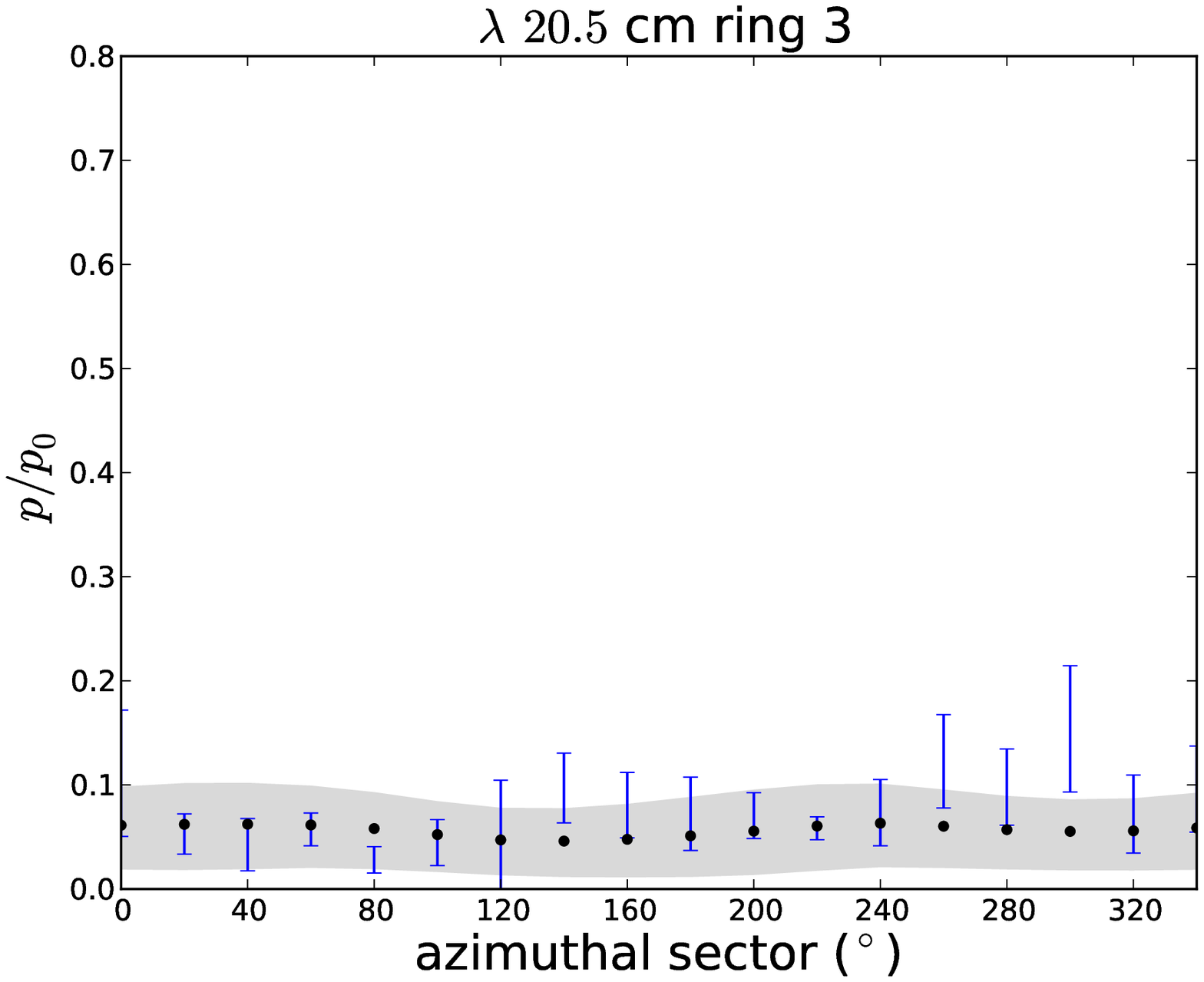} \\
\includegraphics[width=.30\hsize]{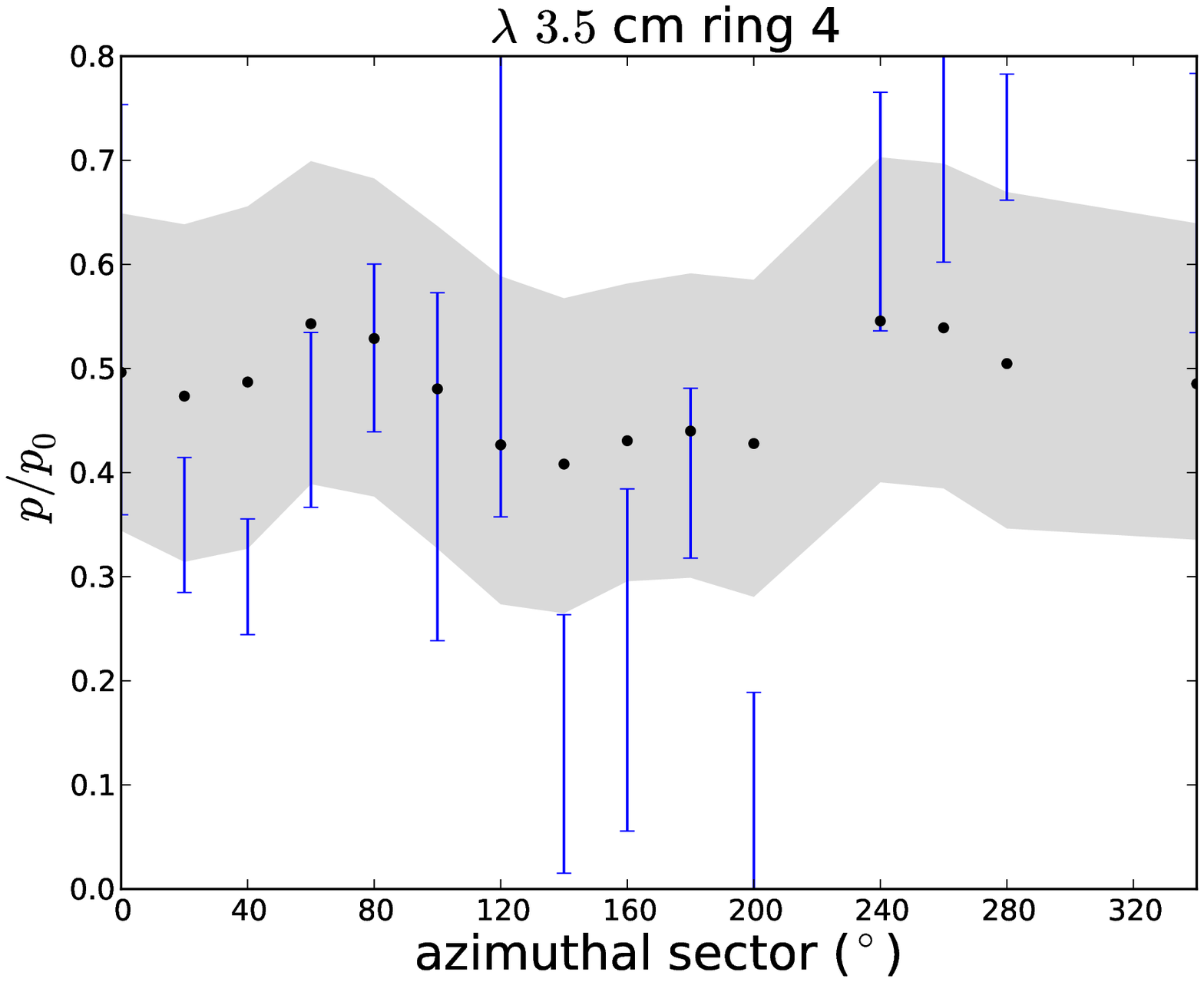}~
\includegraphics[width=.30\hsize]{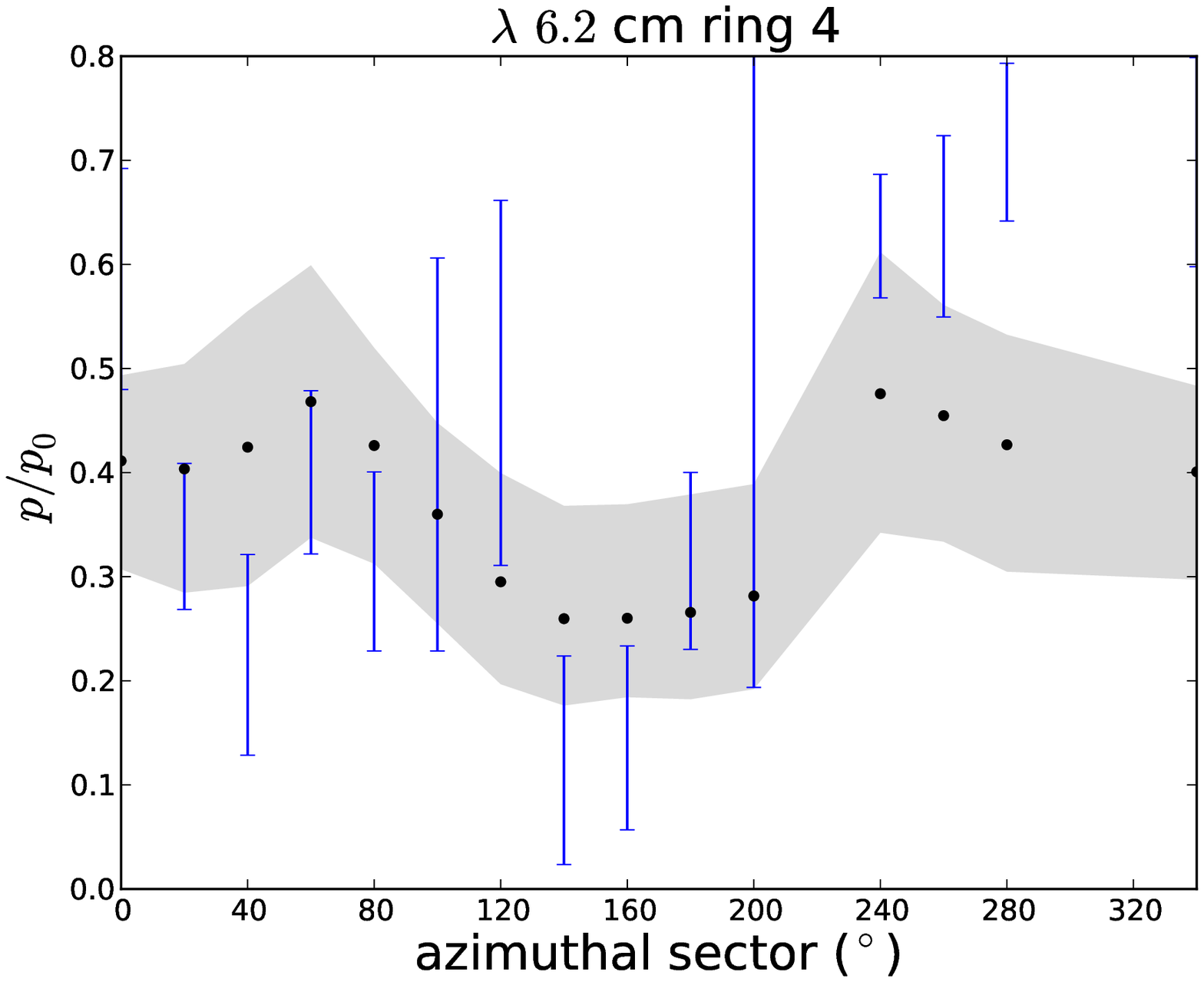}~
\includegraphics[width=.30\hsize]{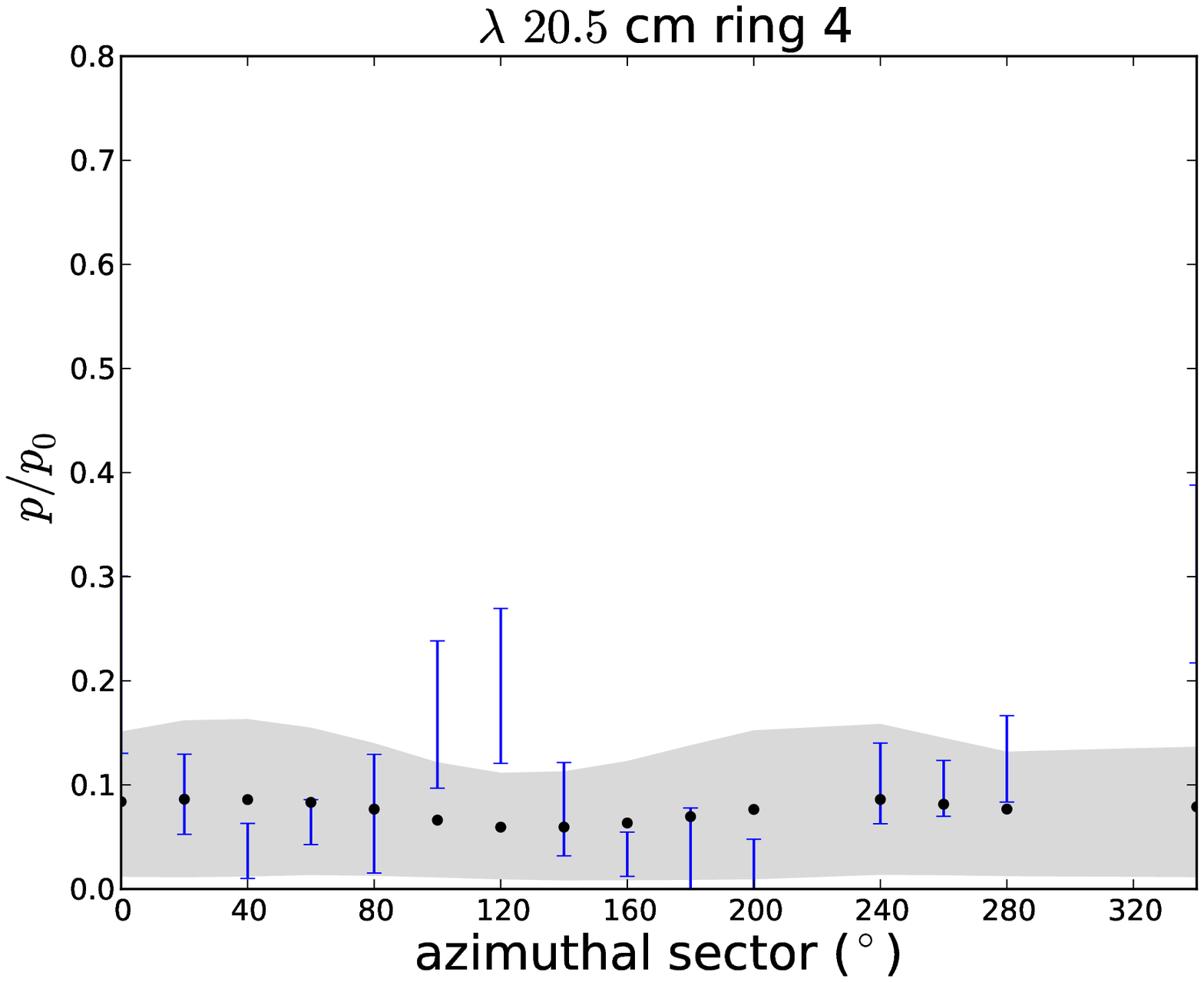} 
\caption{Normalized polarization degree $p/p_0$ as a function of azimuthal angle for observing wavelength of $\lambda \lambda \lambda \, 3.5, 6.2, 20.5$ cm for each of the four rings for a two-layer model. 
Columns provide the polarization profiles per ring at a fixed observing wavelength while rows provide polarization profiles at all three observing wavelengths at a fixed ring.
$0^\circ$ corresponds to the North major axis of M51 with sectors counted counterclockwise.
The solid black points correspond to the predicted polarization value, at each azimuth, from the best-fit magnetic field strengths. The shaded gray region corresponds to the range of polarization values predicted by all regular disk and halo magnetic field configurations encompassed by the solid, $1.5*\chi^{2}_{min}$ contour in Fig.~\ref{Chi_ring1_2_3_4} for rings 1,3,4 and in Fig.~\ref{chi_ring2_revised} for ring 2.
The turbulent magnetic fields are the same as described in Table~\ref{two_layer_strengths_cell_sizes}. The following sectors have been discarded as they are outliers (see text): sector at $60^\circ$ for the inner two rings, and sectors at $220^\circ$, $300^\circ$, and $320^\circ$ in the outermost ring.}
\label{p_profiles_ring1_2_3_4}
\end{figure*}

The lower $\chi^{2}_{min}$ value and more sensitive $\chi^{2}_{red}$ range in ring 1 suggest that the regular and turbulent magnetic fields may be best fit in ring 1 of the two-layer model. This may arise from different magnetic field strengths and thermal electron densities between arm and interarm regions. Ring 1 contains mostly spiral arms, while rings 2 - 4 trace both arm and interarm regions which makes a single fit for magnetic field strengths in the entire ring less of a good fit.
A $\pi$-periodic modulation is apparent in the best-fit polarization profiles of all rings in Fig.~\ref{p_profiles_ring1_2_3_4}, indicating depolarization caused by the regular, mostly azimuthal, magnetic field component. 
It can also be clearly seen that smaller errors in the observed $p/p_0$ decrease the width of the shaded gray corridor.

A model with \emph{only} regular fields does not yield any good fits as expected on physical and observational grounds. A one-layer model is excluded by our modeling as a non-zero regular magnetic field in the halo is predicted by all magnetic field configurations sampled.
This is consistent with the expectation of two separate Faraday rotating layers \citep{berk,fletcher11}. 
We also consider observations of M51 at $610$ MHz which show that $p/p_0 < 1\%$ in spiral arms \citep{farnes_new}. Applying the criterion that $p/p_0 < 1\%$ in the bins that contain the spiral arms in each ring, results in the exclusion of all field configurations which do not have a turbulent magnetic field in the halo. This also automatically rejects a one-layer model.

\subsection{Three-layer model} 

\begin{figure*}	
\centering
\includegraphics[width=.45\hsize]{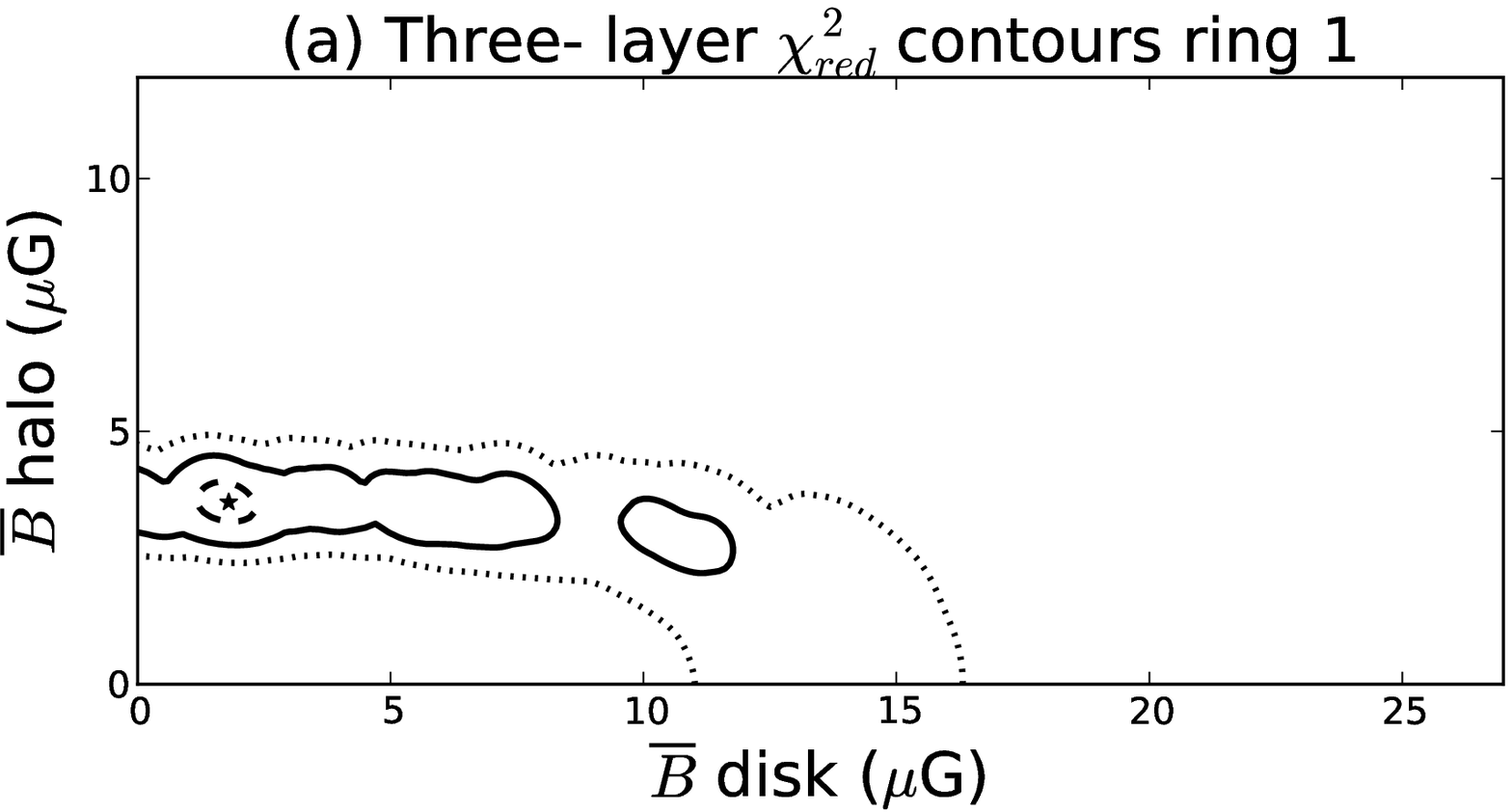}~~~~~~
\includegraphics[width=.45\hsize]{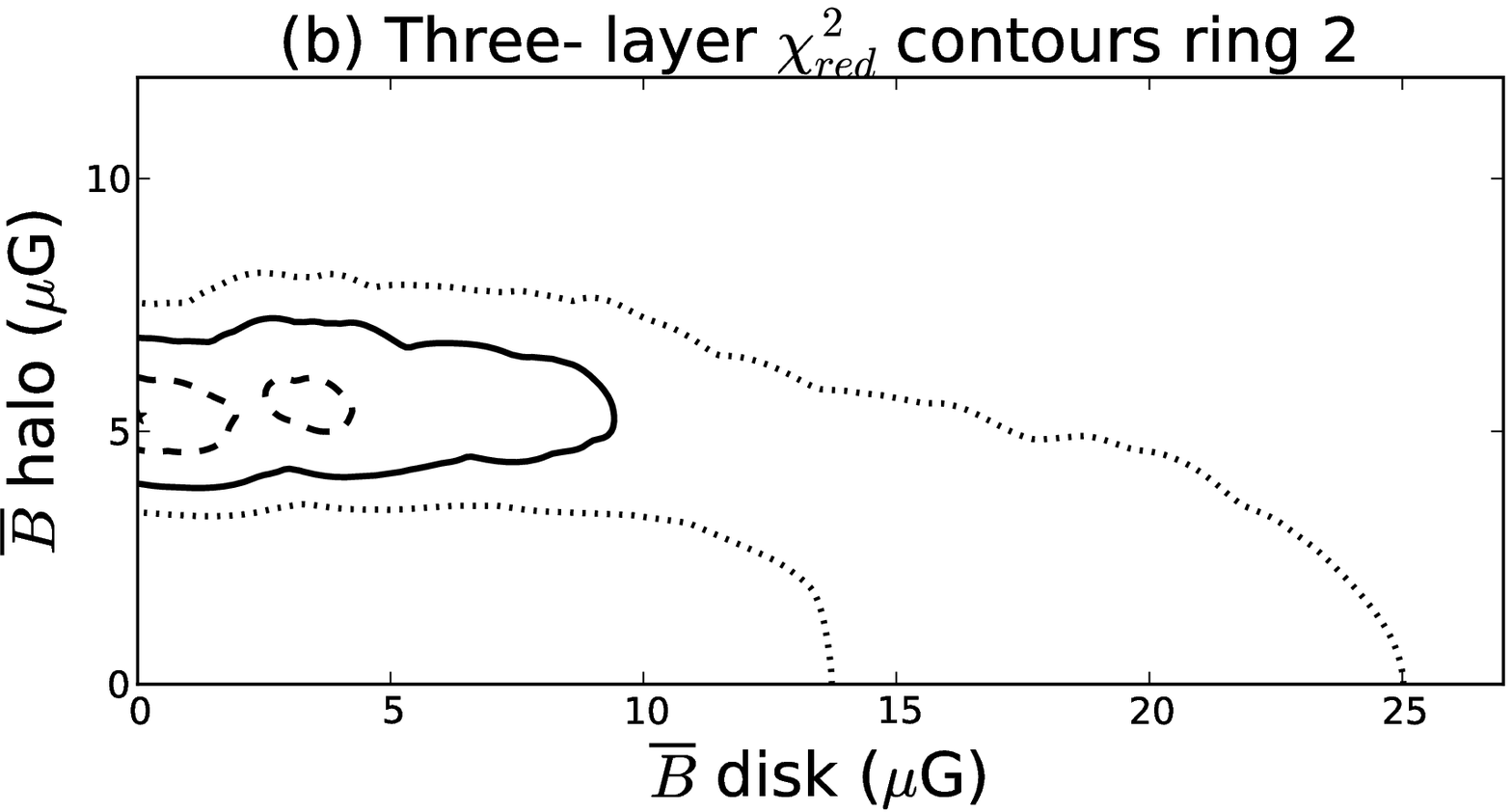} \vspace{1.5cm} \\ 
\includegraphics[width=.45\hsize]{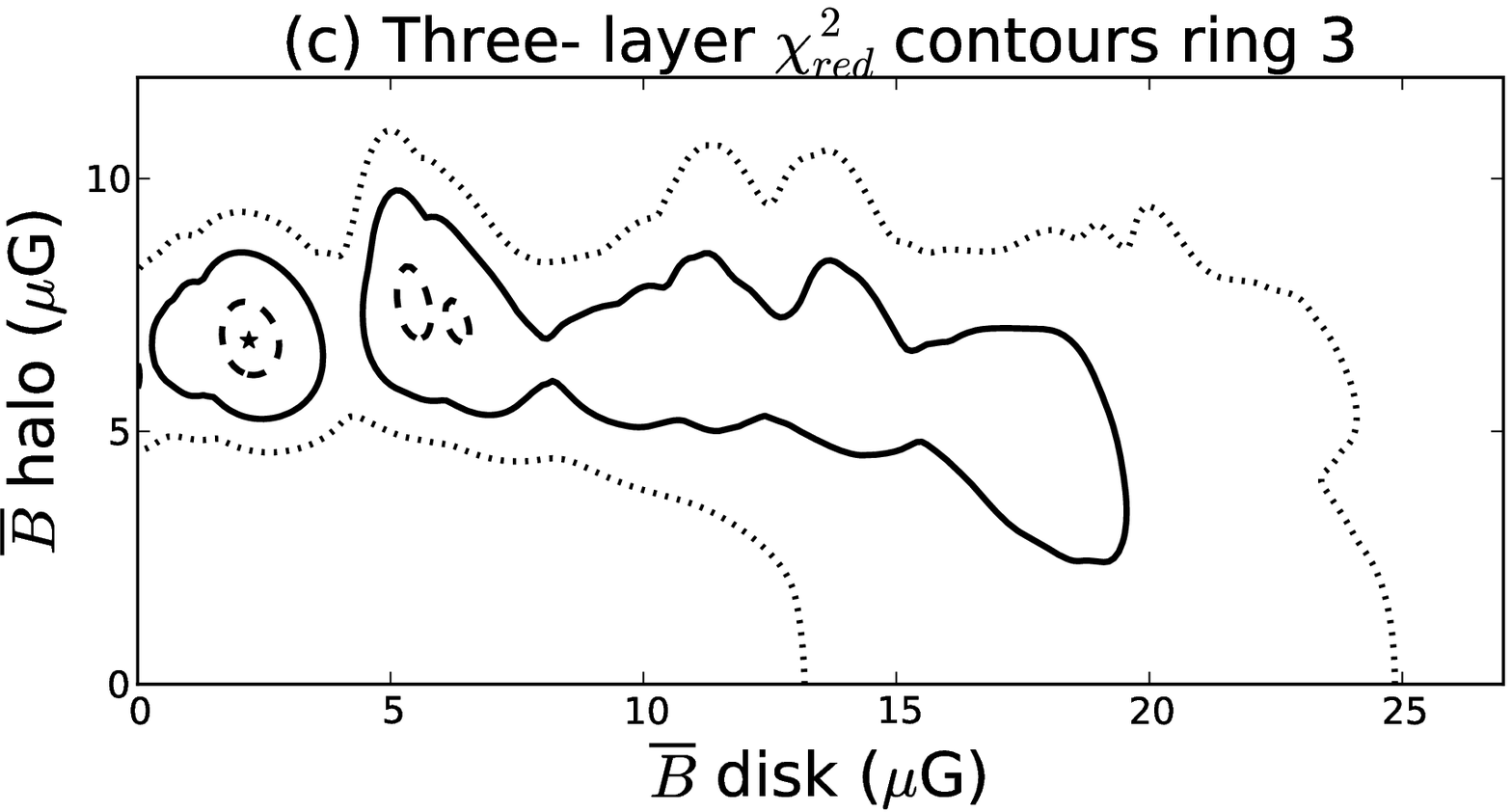}~~~~~~ 
\includegraphics[width=.45\hsize]{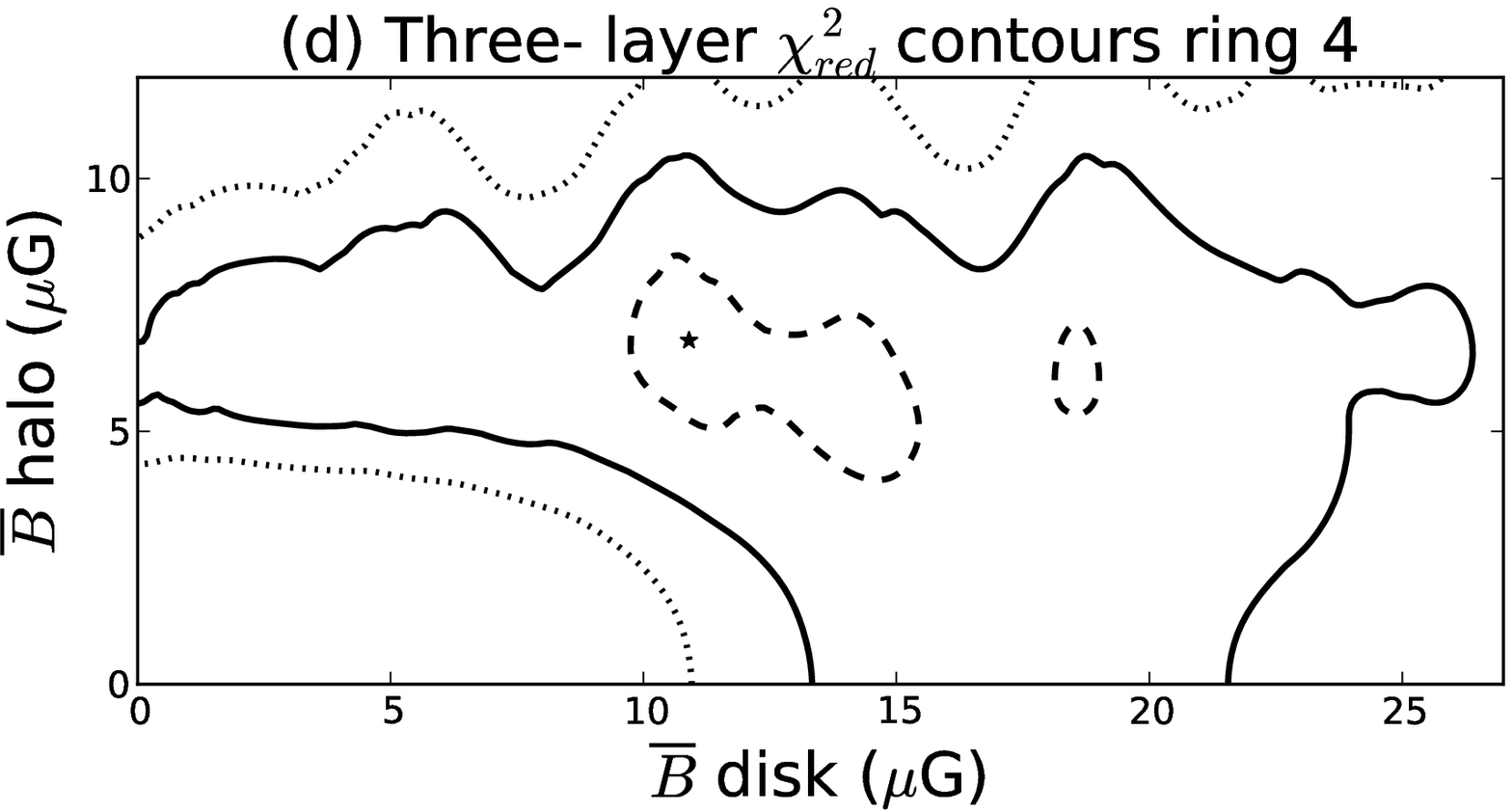}
\caption{(a)-(d) Same as in Fig.~\ref{Chi_ring1_2_3_4} but now for a three-layer DAIHI model.}
\label{Chi_three_layer} 
\end{figure*}

For a three-layer model, with identical near and far sides of the halo, the $\chi^{2}_{red}$ landscape consists of an archipelago of minimum $\chi^{2}_{red}$ values as shown in Fig.~\ref{Chi_three_layer}. 
If a minimum $\chi^{2}_{red}$ were to be taken as representative of a global minimum, then, for the purposes of comparison with the two-layer model, we present the best-fit three-layer model results per ring in Table~\ref{three_layer_strengths_cell_sizes}. 
The three-layer best-fit models are poorer fits to the polarization observations than the two-layer models owing to the higher $\chi^{2}_{min}$ in the innermost pair of rings and the outermost ring.  
Both three- and two-layer models favor the absence of an anisotropic turbulent halo field in all rings. 
Summarizing, the three-layer models result in roughly the same magnetic field values as the two-layer models.

\begin{table}
\caption{Three-layer best-fit DAIHI model magnetic field strengths.}	
\label{three_layer_strengths_cell_sizes} 
\begin{center}
\begin{tabular}{ l  c  c  c  c }
\hline\hline
\rule{0pt}{2.5ex} 
& Ring 1 & Ring 2 & Ring 3 & Ring 4 \\ 
\hline
\rule{0pt}{2.5ex}
Disk &  &  &  & \\
\hline
\rule{0pt}{2.5ex}
Iso.[$\mu$G] 	& $10$ 	& $10$ 	& $10$ 	& $10$  \\
Aniso.[$\mu$G]	& $10$	& $10$	& $10$	& $5$	\\
Reg.[$\mu$G]\tablefootmark{a} 	& ${1.8}^{+10}_{-2}$	& ${0.0}^{+10}_{-0}$	& ${2.2}^{+17}_{-3}$	&	${10.9}^{+16}_{-11}$ \\
d[pc] 		& $40$ 	& $40$ 	& $52$ 	& $61$ \\
\hline
\rule{0pt}{2.5ex}
Halo &  &  &  & \\
\hline	
\rule{0pt}{2.5ex}
Iso.[$\mu$G] 	& $5$	& $8$	& $10$	& $8$	\\
Aniso.[$\mu$G]	& $0$	& $0$	& $0$	& $0$	\\
Reg.[$\mu$G]\tablefootmark{a} 	& $3.6 \pm 1$	& ${5.3}^{+2}_{-1}$	& ${6.8}^{+3}_{-5}$	&	${6.8}^{+4}_{-7}$ \\
d[pc] & $215$ & $157$ & $218$ & $253$ \\
\hline
\rule{0pt}{2.5ex}
$\chi^{2}_{min}$ & $3.0$ & $3.6$ & $2.1$ & $3.6$ \\ 
\hline
\end{tabular} 
\tablefoot{
\tablefoottext{a}{A value of $0 \, \mu$G is to be used for the lower regular field strength bound when the lower error bound exceeds the actual regular field value.} 
}
\end{center}
\end{table}

\subsection{Robustness of results} \label{robustness_results}

The stability of the lowest $\chi^{2}_{red}$ contours for the two-layer models and the (quasi) stability of such contours for the three-layer models, following the bootstrap technique discussed in Section~\ref{procedure}, gives confidence as to the robustness of the results. 
In addition, the elongated shape of the $\chi_{\text{red}}^2$ contours in both these figures indicates that the halo is more sensitive to variation in its regular field value and is therefore a stronger depolarizing region than the disk. The models also yield $\chi^{2}_{red}$ contours for the innermost and outermost pair of rings which are morphologically similar among themselves. Morphological similarity between the rings constituting each pair may be expected based on the physical parameters of thermal electron density and path length being equal for each pair as listed in Table~\ref{modelparam}.

An area of very strong polarized intensity observed at $\lambda \lambda \, 3.5, 6.2$ cm in \citet[Fig.~2]{fletcher11} coincides with the ring 1 sectors at $300^\circ$ and $320^\circ$ and plausibly accounts for the underestimated $p$ values at those locations at all observing wavelengths. Moreover, the ring 1 and ring 2 bins at $60^\circ$ along with the ring 4 bin at $320^\circ$ are outliers as a result of an area of sparse data in the same maps and are consequently discarded. The results shown in Tables~\ref{two_layer_strengths_cell_sizes},~\ref{three_layer_strengths_cell_sizes} are obtained from the outlier free data.

Using the innermost ring which traces the data the closest, our models allow considerable variation in the turbulent magnetic field values in the disk, while magnetic field values in the halo are tightly constrained. 
In particular, replacing the best-fit ring 1 configuration in Table~\ref{two_layer_strengths_cell_sizes} with isotropic and anisotropic turbulent disk fields of $20~\mu$G each, while retaining the $5~\mu$G isotropic turbulent halo field, results in less than a $20 \%$ increase in $\chi_{\text{min}}^2$ whereas only changing the isotropic turbulent halo field to $10~\mu$G, while keeping the isotropic and anisotropic turbulent disk fields at $10~\mu$G and $5~\mu$G, respectively, results in more than a $25 \%$ increase in $\chi_{\text{min}}^2$.  
Correspondingly, total turbulent field values of up to $30~\mu$G are allowed in the disk.  
However, \citet{houde13} report an observed value of the total turbulent disk field of $15~\mu$G in M51, so that any models with a total turbulent field greater than $15~\mu$G are excluded observationally. 
Finally, the {\em regular} disk and halo field strengths vary only slightly for all allowed values of turbulent disk and halo fields, indicating that they are robust for all rings.

\section{Discussion} \label{discussion}

The picture that emerges is the following: in the disk, magnetic field strengths are $B_{\text{reg}}\approx 10~\mu$G and $B_{\text{turb}}\approx 11-14~\mu$G, where $B_{\text{turb}}$ includes both the isotropic and anisotropic random components. In the halo, $B_{\text{reg}}\approx 3~\mu$G and $B_{\text{turb}}$ is about equal to $B_{\text{reg}}$ and consists only of an isotropic component; there is no anisotropic random field in the halo. If anisotropy in magnetic field fluctuations is caused mostly by the strong density waves in M51 and shearing flow, the anisotropy would indeed mostly or exclusively occur in the disk. The regular and total magnetic field strengths in the disk are in agreement with equipartition values of $B_{\text{reg}}\approx 8-13~\mu$G and $B_{\text{tot}}\approx 15-25~\mu$G as calculated by \citet{fletcher11}.

In the halo, maximum cell sizes of the turbulence appear to increase towards the outer part of the galaxy (for a two-layer model), whereas the turbulent cell sizes in the disk are approximately equal. The smaller the turbulent field strength, the larger the turbulent cell size for the representative RM dispersion as given by Eq.~\eqref{diam_turb_cell}. 
If the turbulent cell size in the halo were equal for the inner and outer parts of the galaxy, the RM dispersion would decrease towards the outer part of the galaxy, for the values of turbulent magnetic field resulting from the model, which is not observed. 
However, the cell size in the halo is uncertain since Eq.~\eqref{diam_turb_cell} is only valid for $d \ll D$ and $d \ll L$, which might not be the case in the halo.

The field strengths we find are broadly consistent with earlier studies.
\citet{berk} discussed the magnetic fields in M51 in terms of separate disk and halo for the first time. They found a slightly lower regular magnetic field in the disk $B_{reg,disk}\approx 7~\mu$G, constant across the disk. Their (assumed isotropic) turbulent field strength is comparable to our results; they show that for even larger galactocentric radii out to 15~kpc, this turbulent magnetic field is expected to decrease to $\sim9~\mu$G. \citet{fletcher11} finds regular magnetic field strengths in both the disk and halo between roughly $1 - 4 \, \mu$G with a slight increasing trend in disk regular field strength with radius. They ascribed these anomalously low values to ignoring anisotropic random fields in the equipartition estimate for the regular field strength. 
There is still an anomaly in the estimated regular field strengths though since the polarization angle and RM give $1 - 4 \, \mu$G while depolarization and equipartition both give $10 \, \mu$G field strengths. Possible explanations include ignoring the (unknown) filling factor of the thermal electrons in the RM based estimate, correlations in the line-of-sight distributions of $B_{\parallel}$ and $n_e$, and equipartition not holding.

The resulting magnetic field strengths in the two-layer models and the three-layer models are in agreement. In fact, if the best-fit turbulent magnetic field configurations for all rings for the two-layer model were to be used for a three-layer model, then the resulting best-fit regular disk and halo fields would still be described by the three-layer model within the stated error. 
This implies that all of the signal is depolarized from the far side of the halo, at all wavelengths. Our models therefore confirm the conclusions from \citet{Horellou92} and \citet{berk} based on Faraday rotation and polarization angle measurements. Analyzing polarization data of 21 nearby galaxies from the WSRT SINGS survey \citep{Heald09}, \citet{Braun10} concluded from RM Synthesis that M51 shows polarized intensity at Faraday depths $\phi\approx +13$~rad~m$^{-2}$, coming from a region of emissivity located just above the midplane. They also measured Faraday depth components of about $-180$ and $200$~rad~m$^{-2}$, interpreted as emission from the far side of the mid-plane, which is highly Faraday rotated because of its propagation through the midplane. The positive and negative Faraday depth components roughly coincide to the hemispheres of the disk where the an azimuthal magnetic field would point towards or away from the observer. The high Faraday depth components are consistent with our model, assuming the path length and electron density as in Table~\ref{modelparam} and $B_{\parallel} = 10\sin(l)~\mu$G.
The turbulent cell sizes found for the disk agree with the values in \citep{fletcher11,houde13} and the turbulent cell sizes in the halo are characteristic of the typical cell size expected for spiral galaxies of between $100 - 1000$ pc \citep{sokol}.

The expected total magnetic field strength may also be estimated from the interdependence of the magnetic field strength, gas density, and star formation rate (SFR) as suggested by the far-infrared - radio correlation \citep{Niklas_Beck97}.
\citet{Schleicher13} demonstrated that the observed relation between star formation rate and magnetic field strength arises as a result of turbulent magnetic field amplification by turbulent dynamo action, with turbulence driven by supernova (SN) explosions.
The expression they derived, applied at a redshift $z = 0$, is given by 
\begin{equation} \label{Schleicherformula}
B_{\text{tot}} \sim \sqrt{f_{\text{sat}}8\pi} \, \rho^{1/6}_0 \, \left(f_{\text{mas}} \, \epsilon \, E_{\text{SN}} \right)^{1/3} \, \Sigma^{1/3}_{\text{SFR}},
\end{equation}
where $\rho_0 \sim 10^{-24} \, \mbox{g} \, \mbox{cm}^{-3}$ is the typical ISM density, $\Sigma_{\text{SFR}} \sim 0.1 \, M_\odot \, \mbox{kpc}^{-2} \, \mbox{yr}^{-1}$ is a reference SFR per unit area, $f_{\text{sat}} \sim 5\%$ is the expected saturation level for supersonic turbulence or fraction of the turbulent energy averaged over timescales of $\sim 100$ Myr, $f_{\text{mas}} \sim (8\% / M_\odot)$ is the mass fraction of stars yielding core-collapse SNs, $\epsilon \sim 5\%$ is the fraction of SN energy converted to turbulence, and $E_{\text{SN}} \sim 10^{51} \, \mbox{erg}$ is the typical energy released by an SN.     
The $B_{\text{tot}} \propto \Sigma_{\text{SFR}}^{1/3}$ scaling of \citet{Schleicher13} is comparable with the observed relation between equipartition magnetic field strength and star formation rate for spiral galaxies by \citet{Niklas_Beck97}.
We take $\Sigma_{\text{SFR}} = 0.012 \, M_\odot \, \mbox{kpc}^{-2} \, \mbox{yr}^{-1}$ for M51, adopted from Table 3 of \citet{Tabatabaei13}, which gives a total magnetic field strength $B_{\text{tot}} \sim 10 \, \mu \mbox{G}$ via Eq.~\eqref{Schleicherformula}, as an order of magnitude estimate. 
Considering the roughness of the estimates of the parameter values in Eq.~\eqref{Schleicherformula}, $B_{\text{tot}} \sim 10 \, \mu \mbox{G}$ in the disk is consistent with our results.

\section{Conclusion} \label{conclusion}

We have shown that it is possible to use our analytical depolarization models with radio polarimetric observations, consisting of only three observing wavelengths at $\lambda\lambda\lambda \, 3.5,6.2,20.5$ cm, assisted by the criterion found from the $610$ MHz M51 data by \citet{farnes_new}, to constrain both regular and turbulent magnetic field strengths in M51. By numerically simulating differential Faraday rotation (DFR) and internal Faraday dispersion (IFD) as the main wavelength-dependent depolarization mechanisms along with the contribution of isotropic and anisotropic turbulent magnetic fields to wavelength-independent depolarization, we have arrived at estimates for both regular and turbulent magnetic field strengths in the disk and halo consistent with literature, as shown in Table~\ref{two_layer_strengths_cell_sizes}.

This agreement with earlier studies gives confidence that these models are realistic.
However, our model is more sophisticated than earlier work since it directly simulates the wavelength-dependent depolarizing mechanisms of DFR and IFD thanks to the presence of both regular and random magnetic fields. Previous models \citep{berk,fletcher11} did not include synchrotron emission from the halo, relied primarily on rotation measure (RM) measurements, and did not model the actual contribution of isotropic and anisotropic turbulent magnetic fields to wavelength-independent depolarization.

We find that anisotropic turbulent magnetic field strengths in the disk of M51 are comparable to isotropic turbulent field and regular field strengths ($B \approx 10~\mu$G). However, no anisotropic turbulent field is detected in the halo, where the isotropic field is $B\approx 2~\mu$G, comparable to the regular field strength in the halo.

Comparison of disk-halo models including and excluding a (depolarizing) halo at the far side shows that the far side halo is mostly depolarized at our radio wavelengths, making a two-layer model of disk and near side halo a good approximation.

These models show that even with observational data at only three wavelengths, useful results on magnetic field strengths and configurations can be obtained. Current observational capabilities of broadband radio polarimetry would allow the data to be constrained to a greater extent. 
This would make it possible not only to better determine whether a two-layer or three-layer modeling approach is best suited for describing the data but also to have tighter estimates for the regular and (isotropic and anisotropic) turbulent field strengths in the disk and halo.

Recent studies by \citet{Tabatabaei13} and \citet{Heesen14} have observationally revealed local correlations between the mean and turbulent magnetic field components with the star formation rate with a theoretical motivation for such scenarios recently provided by \citet{Schleicher13}. Future investigations, in conjunction with tests of models for magnetic field amplification by dynamo action, would, therefore, focus on the dynamical physical quantities that give rise to the field structure found in this work. Valuable for this purpose would be spectroscopic data from H$\alpha$ and far-infrared to probe the star formation rate, HI and H$_2$ for estimating gas density, and HI line emission for determination of rotational and turbulent velocity.

\begin{acknowledgements}

  CS and MH acknowledge the support of research program 639.042.915, which is partly financed by the Netherlands Organization for Scientific Research (NWO).
  CS is grateful for the additional financial support by the \emph{Leids Kerkhoven-Bosscha Fonds, LKBF} work visit subsidies.
  AF and AS thank the Leverhulme trust for financial support under grant RPG-097.
  The simulations were performed on the Coma Cluster at Radboud University, Nijmegen, The Netherlands.
  We thank the anonymous referee for the valuable suggestion to connect this work with the broader dynamical picture in galaxies and for suggestions for future research.
\end{acknowledgements}

\bibliographystyle{aa} 

\end{document}